\documentclass[journal]{new-aiaa} 
\usepackage[utf8]{inputenc}

\newcommand{\eg}{{\em e.g.}}
\newcommand{\ie}{{\em i.e.}}

\newcommand{\etal}{{\em et~al.}}

\usepackage{float}
\usepackage{graphicx}
\usepackage{amsmath}

\DeclareMathOperator*{\argmin}{arg\,min}
\usepackage[version=4]{mhchem}
\usepackage{siunitx}
\usepackage{longtable,tabularx}
\usepackage{bm}
\usepackage{todonotes}
\usepackage{algorithm,algpseudocode}
\algnewcommand{\Inputs}[1]{%
  \State \textbf{Inputs:}
  \Statex \hspace*{\algorithmicindent}\parbox[t]{.8\linewidth}{\raggedright #1}
}
\algnewcommand{\Initialize}[1]{%
  \State \textbf{Initialize:}
  \Statex \hspace*{\algorithmicindent}\parbox[t]{.8\linewidth}{\raggedright #1}
}
\algnewcommand{\LineComment}[1]{\State \(\triangleright\) #1}

\newcommand\blfootnote[1]{%
  \begingroup
  \renewcommand\thefootnote{}\footnote{#1}%
  \addtocounter{footnote}{-1}%
  \endgroup
}

\hypersetup{
    colorlinks,
    linkcolor={red!50!black},
    citecolor={blue!50!black},
    urlcolor={blue!80!black}
}

\title{Airfoil Design Parameterization and Optimization using B\'ezier Generative Adversarial Networks}

\author{Wei Chen\footnote{Research Assistant and Ph.D. Candidate, Department of Mechanical Engineering; wchen459@umd.edu.}, Kevin Chiu\footnote{Research Assistant and Ph.D. Student, Department of Mechanical Engineering.} and Mark D. Fuge\footnote{Assistant Professor, Department of Mechanical Engineering.}}
\affil{University of Maryland, College Park, Maryland, 20742}

\begin{document}

\blfootnote{Presented as Paper 2019-2351 at the AIAA Scitech 2019 Forum, San Diego, California, 7-11 January 2019.}

\maketitle

\begin{abstract}
Global optimization of aerodynamic shapes usually requires a large number of expensive computational fluid dynamics simulations because of the high dimensionality of the design space. One approach to combat this problem is to reduce the design space dimension by obtaining a new representation. This requires a parametric function that compactly and sufficiently describes useful variation in shapes. We propose a deep generative model, B\'ezier-GAN, to parameterize aerodynamic designs by learning from shape variations in an existing database. The resulted new parameterization can accelerate design optimization convergence by improving the representation compactness while maintaining sufficient representation capacity. We use the airfoil design as an example to demonstrate the idea and analyze B\'ezier-GAN's representation capacity and compactness. Results show that B\'ezier-GAN both (1)~learns smooth and realistic shape representations for a wide range of airfoils and (2)~empirically accelerates optimization convergence by at least two times compared to state-of-the-art parameterization methods.
\end{abstract}

\section*{Nomenclature}

{\renewcommand\arraystretch{1.0}
\noindent\begin{longtable*}{@{}l @{\quad=\quad} l@{}}
$\bm{c}$  & latent codes \\
$\bm{c}^\dagger$  & near-optimal latent code solution to the first-stage optimization \\
$\bm{c}^*$  & optimal latent code \\
$C_D$  & drag coefficient \\
$C_L$  & lift coefficient \\
$d$  & latent dimension \\
$d'$  & noise dimension \\
$D$  & discriminator \\
$G$  & generator \\
$Ma$  & Mach number \\
$n$  & B\'ezier degree \\
$P$  & control points of the rational B\'ezier curve \\
$P_{\bm{c}}$  & prior distribution of latent codes \\
$P_{data}$  & data distribution \\
$P_G$  & generative distribution \\
$P_{\bm{z}}$  & prior distribution of noise variables \\
$Q$  & auxiliary distribution \\
$Re$  & Reynolds number \\
$t$  & parameter variables of the rational B\'ezier curve \\
$w$  & weights of the rational B\'ezier curve \\
$\bm{x}$  & design variables \\
$\bm{x}^*$  & optimal design \\
$\bm{z}$  & noise variables \\
$\bm{z}^*$  & optimal noise variables \\
$\alpha$  & angle of attack \\
$\lambda_0$,$\lambda_1$,$\lambda_2$,$\lambda_3$,$\lambda_4$  & regularization coefficients of B\'ezier-GAN \\
\end{longtable*}}

\section{Introduction}

\lettrine{A}{erodynamic} shape optimization is a necessary step in designing parts like aircraft wings and (propeller/rotor/turbine) blades. The bottleneck for most high-fidelity aerodynamic shape optimization lies in performing computational fluid dynamics (CFD) simulations given its high computational cost. To perform global optimization, where we search the design space for global optima, the demand for CFD evaluations increases with the number of design variables (\ie, design space dimensionality). While current state-of-the-art can leverage gradient-based techniques, such as adjoint methods, to compute shape derivatives and efficiently guide the search, such approaches are not always feasible\textemdash for example, in chaotic flows where adjoints may become numerically ill-conditioned~\cite{larsson2014prospect} or for solvers that do not calculate adjoints. In such cases, practitioners turn to gradient-free methods. However, in gradient-free methods the need for CFD evaluations can increase exponentially with the design space dimensionality due to the curse of dimensionality~\cite{bellman1957dynamic, regier2015mini}. Even with advanced techniques to balance the exploration and exploitation of the design space~\cite{koziel2013multi, anderson2015aerodynamic, fusi2015drag, amrit2016efficient,masters2017geometric}, this can still make optimization intractable for complex geometries.

To combat this curse of dimensionality, previous research has developed dimensionality reduction (DR) techniques to more compactly represent the original design space. This accelerates exploration by capturing only those dimensions that either affect the final design's performance~\cite{lukaczyk2014active,berguin2014dimensional,berguin2014dimensionality,berguin2015method,grey2018active} or capture major shape variability~\cite{viswanath2011dimension,viswanath2014constrained,cinquegrana2017efficient,cinquegrana2018investigation,yasong2018global,li2018data,poole2019efficient, chen2019aerodynamic}. However, these DR models may not sufficiently or compactly capture the true variation that we observe in real-world designs \textemdash that is, they may requre more dimensions than are needed. Also, the distribution of the reduced design variables corresponding to valid designs is usually unknown, making it difficult to efficiently explore or bound this reduced space. Meanwhile, machine learning researchers have conducted a vast amount of DR research using deep neural networks such as variational autoencoders (VAEs)~\cite{kingma2013auto} and generative adversarial networks (GANs)~\cite{NIPS2014_5423} to learn low-dimensional representations for data from complex high-dimensional distributions. Compared to traditional DR methods, these deep learning frameworks can learn an arbitrary design data distribution parameterized by neural networks. As a result, they can model the highly complex variability of designs using a compact set of variables with a known prior distribution.

In this paper, we apply GANs to learn an interpretable low-dimensional space (\ie, the \textit{latent space}) that encodes major shape variation of aerodynamic designs. Variations not captured by the latent space (\ie, minor features) can be encoded via a \textit{noise space}. The separation of major and minor features allows fast design space exploration while maintaining the representation capacity (\ie, the geometry information is preserved when reducing the design's dimensionality). However, one cannot direct adopt standard GANs to synthesize aerodynamic designs, since they are designed for generating discrete representations such as images and text and do not conserve continuity properties important for aerodynamic shapes. Therefore, we extend the concept of GANs and propose the B\'ezier-GAN for synthesizing smooth aerodynamic designs.

The specific scientific contributions of this paper are: 
\begin{enumerate}
    \item A new type of generative model, B\'ezier-GAN, appropriate for smooth geometry (such as those expressed via splines or B\'ezier curves) that improves the design synthesis quality and convergence rate compared to traditional GANs. It also enables a two-level shape parameterization that separately controls the major and the minor shape deformation. 
    \item A two-stage optimization method that accelerates convergence by prioritizing the optimization of major shape features.
    \item A study of the comparative optima and convergence rate while using several competing parameterization methods \textemdash~our B\'ezier-GAN with different configurations, genetic modal design variables (GMDV)~\cite{kedward2020towards}, singular value decomposition (SVD)~\cite{poole2019efficient}, and B\'ezier surface free-form deformation (FFD)~\cite{sederberg1986free,kenway2017buffet}. We discuss the representation capacity and compactness of each parameterization.
\end{enumerate}

\section{Background}

In this section, we introduce previous work on common algorithms used in aerodynamic design optimization (Sec.~\ref{sec:opt_methods}), parameterization techniques (Sec.~\ref{sec:parameterization}), and methods for reducing design space dimensionality (Sec.~\ref{sec:dr}).

\subsection{Aerodynamic Design Optimization}
\label{sec:opt_methods}

Aerodynamic design is, in large part, an optimization problem. One prototypical example is to find design variables that minimize the drag coefficient $C_D$, while maximizing or constraining the lift coefficient $C_L$~\cite{berguin2014dimensionality,viswanath2014constrained,grey2018active}. In this section, we divide the optimization approaches into two categories, namely gradient-based and gradient-free methods, and introduce the current state-of-the-art in each category.

\subsubsection{Gradient-Based Methods}

Gradient-based methods search for the optimum based on the gradient of the objective function. When the objective is based on CFD simulations, automatic differentiation (AD) or adjoint methods can be used to compute the gradients. Such methods provide a relatively fast and exact method of calculating numerical gradients whose computation cost can be independent of the number of design variables and within an order of magnitude of the forward simulation pass. Because of this, previous work~\cite{mani2012unsteady, kenway2013scalable, tesfahunegn2015surrogate, schramm2016slat, dhert2016rans, wang2018isogeometric, schramm2018turbulent} has used such methods for gradient calculations. Combined with optimization algorithms such as sequential quadratic programming (SQP)~\cite{dhert2016rans, wang2018isogeometric}, steepest descent~\cite{schramm2016slat}, and Newton and quasi-Newton methods~\cite{mani2012unsteady, kenway2013scalable, schramm2018turbulent}, such methods can drastically accelerate gradient calculations in the optimization process~\cite{mani2012unsteady, wang2018isogeometric, schramm2018turbulent}. 

However, adjoint methods can become numerically ill-posed when using simulation techniques like Large Eddy Simulation~\cite{larsson2014prospect}. In addition, AD or adjoint methods require additional simulations beyond the original forward pass to calculate the gradient at each design; depending on the implementation, this can add significant computational costs. In terms of memory, AD or adjoint methods can be expensive compared to, \eg, a finite difference method. Additionally, as a method of gradient calculation, AD or adjoint methods will still inherit some disadvantages of gradient-based algorithms, \eg, converging to local minima in non-convex problems. To mitigate this issue, Berguin \etal~\cite{berguin2015method} use solutions to surrogate-based optimization as starting points for AD methods, hoping to find good local optima.

\subsubsection{Gradient-free Methods}

As mentioned earlier, there are cases where the optimization problem is ill-conditioned or gradients of the objective are inaccessible. Gradient-free methods can side step this problem and can increase the likelihood of finding a global optimum~\cite{venter2003particle}. Two classes of approaches are widely used for gradient-free aerodynamic shape optimization~\textemdash~population-based optimization (PBO) and surrogate-based optimization (SBO).

A popular PBO method in aerodynamic shape optimization is the genetic algorithm (GA)~\cite{viswanath2011dimension,cinquegrana2018investigation,jeong2018turbine,skinner2018review}. It mimics the process of biological evolution through mutation, recombination, and reproduction of different designs. Work has also been done to augment GA with the bees algorithm~\cite{tandis2017bees} and adaptive mutation rates~\cite{jahangirian2017adaptive}, resulting in more accurate optimization and/or faster convergence. Other PBO methods applied in aerodynamic optimization are differential evolution~\cite{liu2018weed} and particle swarm optimization~\cite{venter2003particle,ray2004swarm}. However, due to the large number of CFD evaluations needed to form the population at each iteration, PBO methods can usually be prohibitively expensive computationally, especially if every evaluation requires a high-fidelity CFD simulation~\cite{skinner2018review}.

Surrogate-based optimization uses an inexpensive surrogate model to approximate the expensive CFD evaluations. Bayesian optimization (BO) is a commonly used SBO method. It consists of two components: a sampling strategy (\eg, maximum expected improvement~\cite{jones1998efficient} or maximum upper confidence bound~\cite{srinivas2009gaussian}) and a surrogate modeling method (\eg, Gaussian process regression, also known as kriging~\cite{rasmussen2004gaussian}). In each iteration, the sampling strategy proposes a point in the design space for evaluation, which is then used to update the surrogate model. Compared to methods like GAs, surrogate-based optimization reduces the number of expensive CFD evaluations needed in aerodynamic shape optimization~\cite{viswanath2011dimension,lukaczyk2014active,berguin2014dimensionality,cinquegrana2017efficient,han2018aerodynamic}. However, for a high-dimensional design space, the number of evaluations will still be inevitably high due to the curse of dimensionality. This creates a demand for shape parameterization methods or design space dimensionality reduction techniques that can reduce the number of design variables in optimization.

There are other gradient-free methods beyond the above two classes (\eg, simulated annealing~\cite{wang2001aerodynamic, manzo2008drawing}) but are not as commonly used in aerodynamic shape optimization. We direct interested readers to Ref.~\cite{skinner2018review} for a thorough review of these methods.

\subsection{Shape Parameterization}
\label{sec:parameterization}

A parameterization maps a set of parameters to points along a smooth curve or surface via a parametric function. Common parameterization for aerodynamic shapes includes splines (\eg, B-spline and B\'ezier curves)~\cite{je2001optimized,venkataraman1995new,rogalsky2000differential}, free-form deformation (FFD)~\cite{sederberg1986free,kenway2017buffet}, class-shape transformations (CST)~\cite{nadarajah2007survey,kulfan2008universal}, PARSEC~\cite{li1998manual,sobieczky1999parametric}, and B\'ezier-PARSEC~\cite{derksen2010bezier}. Usually during design optimization, parameters are sampled to generate design candidates~\cite{han2018aerodynamic}. There are two main issues when optimizing these parameters from conventional parameterization: (1)~one has to guess the limits of the parameters to form a bounding-box within which the optimization operates, and (2)~the design space dimensionality is usually higher than the underlying dimensionality for representing sufficient shape variability~\cite{chen2017design}~\textemdash \ie, to capture sufficient shape variation, manually designed shape parameterizations require higher dimensions than are strictly necessary. In contrast, this paper shows that learned parameterizations can often achieve much higher compactness without a loss in representation quality or optimization performance.

\subsection{Design Space Dimensionality Reduction}
\label{sec:dr}

It is usually wasteful to search for optima in the spaces of aforementioned shape parameters, since valid designs only constitute a small portion of those spaces so that most CFD evaluations are performed on invalid designs. Past work has studied methods to obtain more compact representations via dimensionality reduction. Factor screening methods~\cite{welch1992screening,myers1995response} can select the most relevant design variables for a design problem while fixing the rest as constant during optimization. These methods fail to consider the correlation between design variables. In response, researchers have studied ways to capture the low-dimensional subspace that identifies important directions with respect to the change of \textit{response} (\ie, QoI or performance measure)~\cite{lukaczyk2014active,berguin2014dimensional,berguin2014dimensionality,berguin2015method,grey2018active}. This response-based dimensionality reduction usually has several issues: 
(1)~it requires many simulations when collecting samples of response gradients;
(2)~variation in gradients can only capture non-linearity rather than variability in the response, so extra heuristics are required to select latent dimensions that capture steep linear response changes;
(3)~the learned latent space is not reusable for any different design space exploration or optimization task (\ie, when a different response is used); and 
(4)~the linear DR techniques applied in previous work may not model well responses with non-linear correlation between partial derivatives.

The first three issues can be avoided by directly applying DR on design variables without associating them with the response. Doing so assumes that if changes in a design are negligible, changes in the responses are also negligible. In the area of aerodynamic design, researchers use linear models such as principal component analysis (PCA)~\cite{cinquegrana2017efficient,cinquegrana2018investigation,yasong2018global} or singular value decomposition (SVD)~\cite{li2018data,allen2018wing,poole2019efficient} to reduce the dimensionality of design variables. 
Although those linear models provide optimal solutions to the linear DR problem, their linear nature makes them unable to achieve the most compact representation (\ie, use the least dimensions to retain similar variance in the data) when the data is nonlinear, which is the most common case for real-world data.
Nonlinear models like generative topographic mapping~\cite{viswanath2011dimension,viswanath2014constrained} can solve this problem to some extent, but are still limited to the assumption that data follow a Gaussian mixture distribution, which is too strict in most real-world cases. Beyond these data-driven methods, genetic modal design variables (GMDV)~\cite{kedward2020towards} generates airfoils through orthogonal modes derived from the reduced singular matrix of the third-difference matrix. None of these DR methods encourage compactness of the reduced shape representation, where the volume of the latent space that maps to the domain of invalid designs are minimized.
Complementary work in DR has been done in other fields such as computer vision and computer graphics~\cite{lee2007nonlinear,goodfellow2016deep}, where DR is used for generating images or 3D shapes. Deep generative networks such as VAEs and GANs have been widely applied in those areas to learn the latent data representations. These methods are known for their ability to learn a compact latent representation from complex high-dimensional data distributions, where the latent representation follows a simple, known distribution (\eg, a normal or uniform distribution).
Our work extends this class of techniques by considering the generation of smooth geometries such as those needed in spline-based representations.

Note that our assumption is that we already have a reasonably distributed dataset and the target design is within or at least not far from the coverage of the data distribution. This assumption exists for all data-driven methods. For example, given a database of airfoils, a data-driven method may help solve an airfoil design problem, but would not necessarily give a good solution for a hydrofoil design problem.

As DR models map latent variables to shapes, we can treat the latent variables and the mapping as parameters and the parametric function, respectively. Thus, in a broader sense, we will also refer to these methods as parameterizations throughout this paper, though these are distinct from prior work in parameterizations since they are inferred from data directly, rather than fixed apriori.

\section{Learning Compact and Disentangled Representations}

In this section, we introduce two deep generative models~\textemdash~GAN~\cite{NIPS2014_5423} and its variant, InfoGAN~\cite{chen2016infogan}, which our proposed B\'ezier-GAN builds upon. 
Given a set of existing aerodynamic designs (\eg, airfoils from the UIUC database), a deep generative model like GAN can learn a mapping from a known distribution to the unknown distribution of existing designs. We call the set of existing designs the \textit{training data}. In GANs, we usually set the known distribution as a normal or uniform distribution. Samples drawn from this distribution are called \textit{noise}. The noise dimension is typically much lower than the design space dimension and hence we can treat the noise as a reduced representation of designs.

\begin{figure}
\centering
\includegraphics[width=0.8\textwidth]{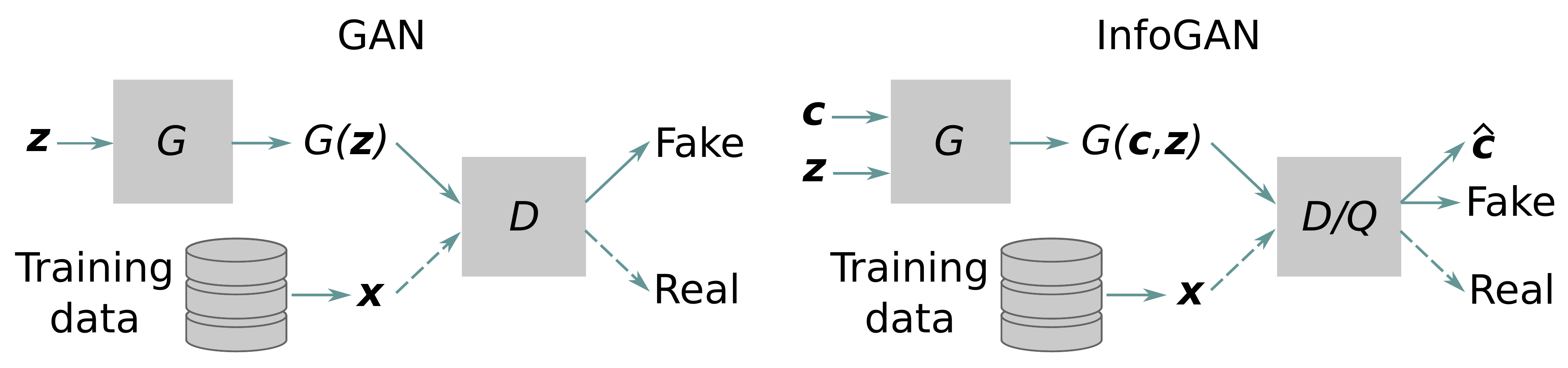}
\caption{Architectures of GAN and InfoGAN.}
\label{fig:gan}
\end{figure}

A GAN consists of two components: a generator $G$ and a discriminator $D$ (Fig.~\ref{fig:gan}). The generator takes in random noise $\bm{z}$ from some known prior distribution $P_{\bm{z}}$ and generates data $\Tilde{\bm{x}}=G(\bm{z})$. The discriminator takes in a sample (either from the training data or synthesized by the generator) and predicts the probability of the sample coming from the training data. The generator tries to make the generative distribution $P_G$ look like the data distribution $P_{data}$ to fool the discriminator; the discriminator tries not to be fooled. GANs achieve this by solving the following minimax problem:
\begin{equation}
\min_G\max_D V(D,G) = \mathbb{E}_{\bm{x}\sim P_{data}}[\log D(\bm{x})] + \mathbb{E}_{\bm{z}\sim P_{\bm{z}}}[\log(1-D(G(\bm{z})))].
\label{eq:gan}
\end{equation}
Both $D$ and $G$ components improve via training until the discriminator cannot differentiate between real (data) and fake (synthetic) inputs, implying that the generative distribution resembles the data distribution. We direct interested readers to~\cite{NIPS2014_5423} for a more detailed explanation of GANs.

Standard GANs do not have a way of regularizing the latent representation (noise); thus, the noise may end up being non-interpretable. This causes the noise variation not to reflect an intuitive design variation, which impedes design space exploration. To compensate for this weakness, the InfoGAN encourages interpretable and disentangled latent representations by maximizing the mutual information between the \textit{latent codes} $\bm{c}$ and the generated samples $\Tilde{\bm{x}}$. Thus, InfoGAN's generator takes $\bm{c}$ as an additional input, \ie, $\Tilde{\bm{x}}=G(\bm{c},\bm{z})$ (Fig.~\ref{fig:gan}). Unfortunately, it is hard to directly maximize the mutual information $I(\bm{c}; G(\bm{c},\bm{z}))$, so instead, an InfoGAN approximates the solution by maximizing a lower bound. The mutual information lower bound $L_I$ is
\begin{equation}
L_I(G,Q) = \mathbb{E}_{\bm{x}\sim P_G}[\mathbb{E}_{\bm{c}'\sim P(\bm{c}|\bm{x})}[\log Q(\bm{c}'|\bm{x})]] + H(\bm{c}),
\label{eq:li}
\end{equation}
where $H(\bm{c})$ is the entropy of the latent codes, and $Q$ is the auxiliary distribution for approximating $P(\bm{c}|\bm{x})$. We direct interested readers to~\cite{chen2016infogan} for the derivation of $L_I$. The InfoGAN's loss function combines $L_I$ with the standard GAN's loss:
\begin{equation}
\min_{G,Q}\max_D \mathbb{E}_{\bm{x}\sim P_{data}}[\log D(\bm{x})] + \mathbb{E}_{\bm{c}\sim P_{\bm{c}},\bm{z}\sim P_{\bm{z}}}[\log(1-D(G(\bm{c},\bm{z})))] - \lambda L_I(G,Q),
\label{eq:infogan}
\end{equation}
where $P_{\bm{c}}$ is the prior distribution of latent codes and $\lambda$ is a weight parameter. In practice, $H(\bm{c})$ is treated as constant if the distribution of $\bm{c}$ is fixed. The auxiliary distribution $Q$ is also learned by a neural network and is simply approximated by sharing all the convolutional layers with $D$ and adding an extra fully connected layer to $D$ to predict the conditional distribution $Q(\bm{c}|\bm{x})$. Thus, as shown in Fig.~\ref{fig:gan}, the discriminator tries to predict both the source of the data and the latent codes~\footnote{Here we use the discriminator $D$ to denote both $Q$ and $D$, since they share neural network weights.}. 
Note that although the InfoGAN encourages disentangled latent representation, it does not guarantee orthogonal latent variables as PCA or SVD does. Instead, it guarantees statistical independence among latent or noise variables when each variable is independently sampled. This is similar in concept to what independent component analysis (ICA)~\cite{comon1994independent} does for linear DR. We build upon the work of InfoGAN, extending it to spline-based geometry.

\section{B\'ezier-GAN: Spline-Based Shape Synthesis}

Typical approaches to generative shape models (such as GANs) represent shapes as a collection of discrete samples (\eg, as pixels or voxels) owing to the their original development in the computer vision community. For example, a na\"ive way of synthesizing aerodynamic shapes like airfoils would be to generate this \textit{discrete representation} directly using the generator, such as generating a fixed number of coordinates sampled along the airfoil's surface curve (\eg, Fig.~\ref{fig:iterations}, left).
However, in practice, aerodynamic shapes typically possess substantial smoothness/continuity and are typically represented using parametric curve families like splines, B\'ezier curves, or NURBS surfaces. The na\"ive GAN representation of predicting discretized curves from the generator usually (1) creates noisy curves that have low smoothness and (2) have parametric output that is harder for humans to interpret and use in standard CAD packages compared to equivalent curve representations (\eg, B\'ezier curves). This creates problems, particularly in aerodynamic shape synthesis. 

To solve this issue, we modified the InfoGAN's generator such that it only generates smooth shapes that conform to B\'ezier curves. We call this generative adversarial network a B\'ezier-GAN~\cite{chen2018b}. 

\subsection{Architecture}

\begin{figure}[hbt!]
\centering
\includegraphics[width=1\textwidth]{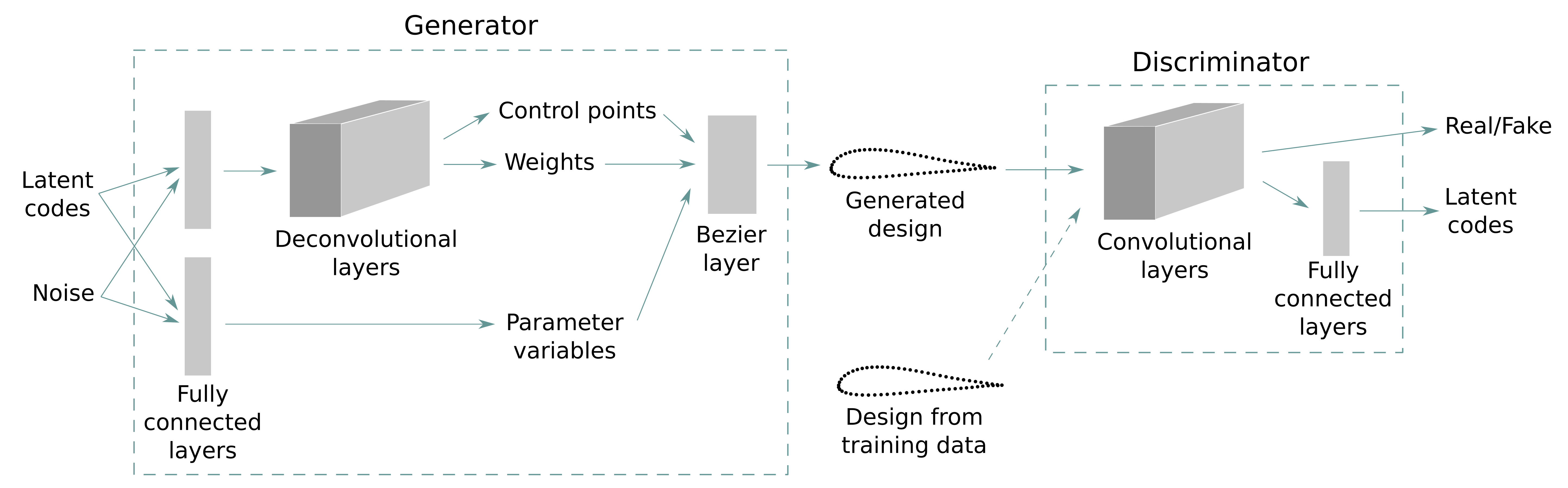}
\caption{Model architecture of the B\'ezier-GAN.}
\label{fig:architecture}
\end{figure}

As shown in Fig.~\ref{fig:architecture}, the overall architecture is adapted from the InfoGAN. However, instead of directly outputting discrete coordinates along the curve, the generator synthesizes \textit{control points} $\{P_i|i=0,...n\}$, \textit{weights} $\{w_i\geq 0|i=0,...n\}$, and \textit{parameter variables} $\{0\leq t_j\leq 1|j=0,...,m\}$ of rational B\'ezier curves, where $n$ is the B\'ezier degree, and the number of surface points to represent the curve is $m+1$. The last layer of the generator\textemdash the B\'ezier layer\textemdash converts this rational B\'ezier curve representation into discrete representation $\bm{x}$:
\begin{equation}
\bm{x}_j = \frac{\sum_{i=0}^n\binom ni t_j^i(1-t_j)^{n-i}P_i w_i}{\sum_{i=0}^n\binom ni t_j^i(1-t_j)^{n-i}w_i},~~~~j=0,...,m.
\label{eq:bezier}
\end{equation}
Since $\bm{x}_j$ is differentiable with respect to $\{P_i\}$, $\{w_i\}$, and $\{t_j\}$, we can train the network using back propagation. To improve numerical stability, the Bernstein polynomial is computed via its natural logarithm:
\begin{equation}
B_i^n(t)=\binom ni t^i(1-t)^{n-i} = \exp\left(\log\Gamma(n+1)-\log\Gamma(i+1)-\log\Gamma(n-i+1)+i\log t+(n-i)\log(1-t)\right),
\end{equation}
where $\Gamma$ denotes the gamma function.

\begin{figure}[hbt!]
\centering
\includegraphics[width=0.3\textwidth]{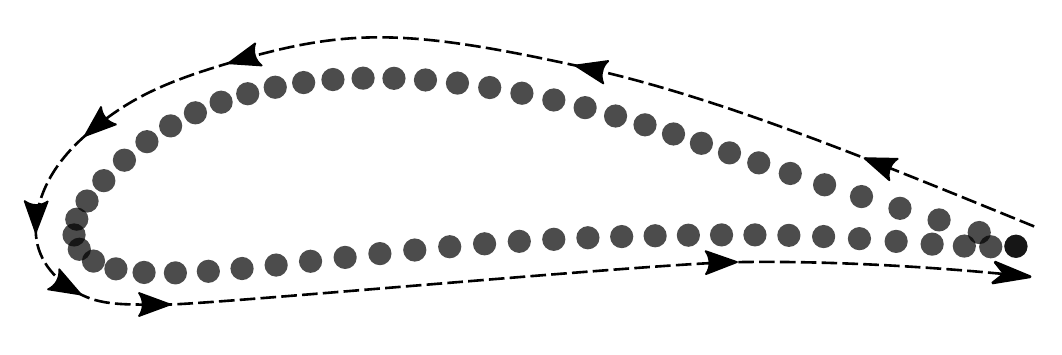}
\caption{Ordering of the airfoil representation.}
\label{fig:order}
\end{figure}

Particularly, as is shown in Fig.~\ref{fig:order}, an aerodynamic surface (here we use airfoil as an example) can be represented as an ordered sequence of points. Thus, the parameter variables have to satisfy $0 = t_0 < t_{j-1} < t_j < t_m = 1, j=2,...,m-1$. Instead of directly generating $\{t_j|j=0,...,m\}$, we use a softmax activation to generate the intervals $\{\delta_j=t_j-t_{j-1}|j=1,...,m\}$ and then compute the cumulative sum of $\{\delta_j\}$ to get $\{t_j\}$. 

\subsection{Regularization}

The rational B\'ezier representation (\ie, the choice of $\bm{P}$, $\bm{w}$, and $\bm{t}$) for a point sequence is not unique. For example, we have observed that the generated control points are dispersed and disorganized. The weights vanish at control points far away from the surface points, and the parameter variables have to become highly non-uniform to adjust the ill-behaved control points. To prevent B\'ezier-GAN from converging to bad optima, we regularize these B\'ezier parameters.

\subsubsection{Control Points}

Since the control points can be dispersed and disorganized, causing the weights and parameter variables to also behave abnormally, one way to regularize control points is to keep them close together. We use the average and maximum Euclidean distance between each two adjacent control points as a regularization term:
\begin{equation}
R_1(G) = \frac{1}{Nn}\sum_{s=1}^N\sum_{i=1}^n \big\|\bm{P}_i^{(s)}-\bm{P}_{i-1}^{(s)}\big\|,
\label{eq:reg_cp}
\end{equation}
where $N$ is the sample size.

\subsubsection{Weights}

To eliminate the effects of redundant control points and avoid convoluted curves, we regularize the weights of control points except for the first and the last ones:
\begin{equation}
R_2(G) = \frac{1}{Nn}\sum_{s=1}^N\sum_{i=1}^{n-1} \big|w_i^{(s)}\big|.
\label{eq:reg_w}
\end{equation}

\subsubsection{Edge Alignment}

We represent the aerodynamic surface curve by 2D points starting from and ending at the trailing edge, as shown in Fig.~\ref{fig:order}. Thus, the first and last control points ($\bm{P}_0$ and $\bm{P}_n$) should meet at the trailing edge so that the surface curve can be closed. To enforce this constraint, we add the following regularization term to the loss function:
\begin{equation}
R_3(G) = \frac{1}{N}\sum_{s=1}^N \big\|\bm{P}_0^{(s)}-\bm{P}_n^{(s)}\big\|.
\label{eq:reg_end1}
\end{equation}

In addition, we use the following regularization term to ensure that the surface curve does not self-intersect near the trailing edge:
\begin{equation}
R_4(G) = \frac{1}{N}\sum_{s=1}^N \max\left\{0, -10\big(\bm{P}_{0y}^{(s)}-\bm{P}_{ny}^{(s)}\big)\right\},
\label{eq:reg_end2}
\end{equation}
where $\bm{P}_{iy}^{(s)}$ denotes the $y$-coordinate of the $i$-th control point in the $j$-th airfoil sample. 

With the above regularization terms, the loss function of B\'ezier-GAN becomes 
\begin{equation}
\min_{G,Q}\max_D V(D,G)-\lambda_0 L_I(G,Q)+\sum_{r=1}^4\lambda_r R_r(G),
\label{eq:loss}
\end{equation}
where $\lambda_i$ controls the weight of each corresponding regularization term.

\section{Two-Stage Optimization over the B\'ezier-GAN Parameterization}

For a normal parameterization $F$ (e.g., FFD), we can synthesize an airfoil design $\bm{x}$ (\ie, the Cartesian coordinates of surface points) through some given parameters $\bm{p}$ (e.g., in FFD, $\bm{p}$ is the coordinates of control points). This synthesis process can be expressed as $\bm{x} = F(\bm{p})$. The CFD simulator is also a function $f$ that maps a design $\bm{x}$ to its performance metric $y$ (\eg, the drag coefficient). We can write this process as $y=f(\bm{x})=f(F(\bm{p}))$. The optimization problem can be expressed as
$$\min_{\bm{p}} y = f(F(\bm{p}))$$

If we use B\'ezier-GAN's generator $G$ as a paramterization, there will be two sets of parameters we must optimize over~\textemdash~the latent code $\bm{c}$ and the noise variables $\bm{z}$. The synthesis process can be written as $\bm{x}=G(\bm{c}, \bm{z})$. Thus, the optimization becomes
$$\min_{\bm{c}, \bm{z}} y = f(G(\bm{c}, \bm{z}))$$
But instead of optimizing $\bm{c}$ and $\bm{z}$ simultaneously, we propose a two-stage optimization (TSO) designed to exploit B\'ezier-GAN's two-level parameterization for faster convergence towards good solutions (Alg.~\ref{alg:tso}). 

\begin{figure}[hbt!]
\centering
\includegraphics[width=0.9\textwidth]{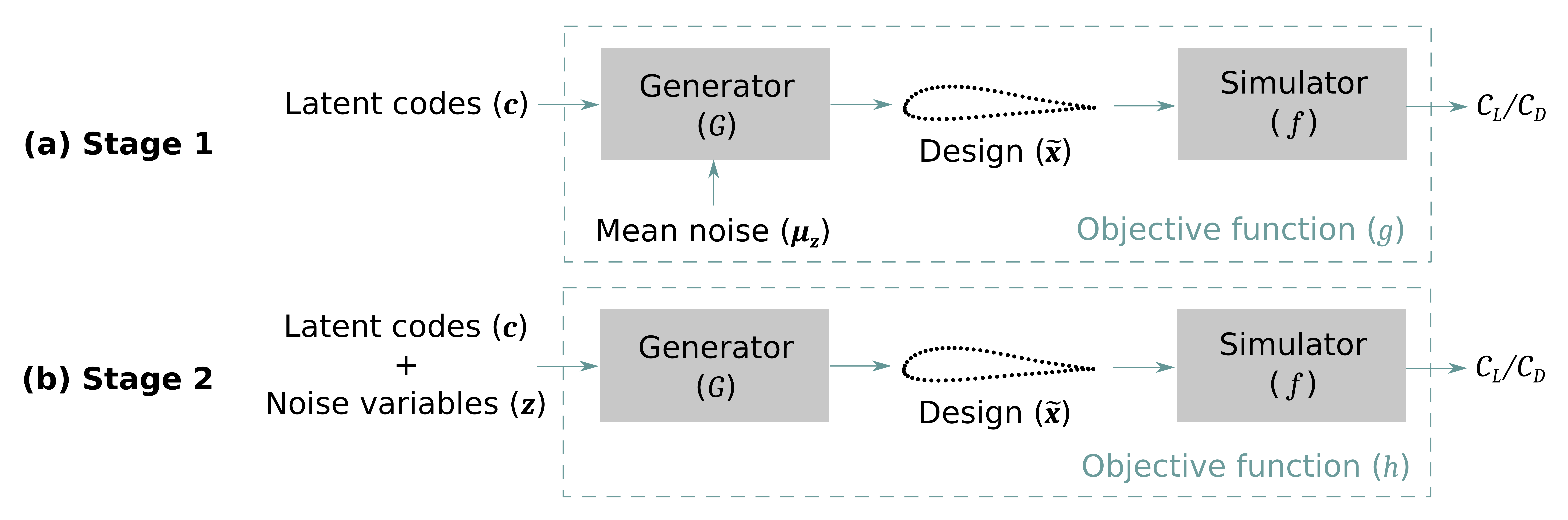}
\caption{Objective functions in both stages of two-stage optimization (TSO).}
\label{fig:tso_obj}
\end{figure}
 
\begin{algorithm}
\caption{Two-stage optimization (TSO)}
\label{alg:tso}
\begin{algorithmic}[1]
    \LineComment{Given the latent dimension $d$, the noise dimension $d'$, the mean noise $\bm{\mu_z}$, a trained generator $G$, a CFD simulator $f$, and evaluation budget $T$}
\Procedure{TSO}{$d, d', G, f, T$}
    \State $T_1 \leftarrow Td/(d+d')$, $T_2 \leftarrow T-T_1$ \Comment{Assign budget to each stage}
    \State For $t = 1:T_1$, solve $\min_{\bm{c}} g(\bm{c})=f(G(\bm{c},\bm{\mu_z}))$ \Comment{First-stage optimization}
    \State $\bm{c}^\dagger \leftarrow \argmin_{\bm{c}} g(\bm{c})$
    \State For $t = 1:T_2$, solve $\min_{\bm{c},\bm{z}} h(\bm{c},\bm{z})=f(G(\bm{c},\bm{z}))$ with $(\bm{c}^\dagger, \bm{\mu_z})$ as a warm start \Comment{Second-stage optimization}
    \State $\bm{c}^*, \bm{z}^* \leftarrow \argmin_{\bm{c},\bm{z}} h(\bm{c},\bm{z})$
    \State $\bm{x}^* \leftarrow G(\bm{c}^*, \bm{z}^*)$ \Comment{Synthesize the optimal design}
	\State\Return $\mathbf{x}^*$
\EndProcedure
\end{algorithmic}
\end{algorithm}

Since we maximize the mutual information of the lower bound between the latent codes and the generated design, the latent codes $\bm{c}$ will capture major shape variations, while minor variations are controlled by the noise $\bm{z}$. This is further demonstrated in Sec.~\ref{sec:experiment}. This two-level representation allows us to perform design space exploration in a more efficient way. Specifically, we first optimize $\bm{c}$ with fixed $\bm{z}$ to quickly find a near-optimal solution $\bm{c}^\dagger$:
\begin{equation}
\bm{c}^\dagger = \argmin_{\bm{c}} g(\bm{c}),
\label{eq:opt_latent}
\end{equation}
where $g(\bm{c})=f(G(\bm{c}, \bm{\mu_z}))$ and $\bm{\mu_z}=\mathbb{E}_{\bm{z}\sim P_{\bm{z}}}[\bm{z}]$ (Fig.~\ref{fig:tso_obj}a). When $P_{\bm{z}}$ is a normal distribution centered at 0, we have $\bm{\mu_z}=\bm{0}$. Then using $(\bm{c}^\dagger, \bm{\mu_z})$ as a warm start, we optimize both $\bm{c}$ and $\bm{z}$ to refine the near-optimal solution:
\begin{equation}
\bm{c}^*, \bm{z}^* = \argmin_{\bm{c},\bm{z}} h(\bm{c},\bm{z}),
\label{eq:opt_overall}
\end{equation}
where $h(\bm{c},\bm{z})=f(G(\bm{c}, \bm{z}))$ (Fig.~\ref{fig:tso_obj}b). We can then synthesize the optimal shape $\bm{x}^*=G(\bm{c}^*, \bm{z}^*)$.

For the \textit{first-stage optimization} (\ie, optimizing $\bm{c}$), we want to quickly find a good solution $\bm{c}^\dagger$ near the true underlying optimum. In this paper we use an SBO method called Efficient Global Optimization (EGO)~\cite{jones1998efficient}. This method minimizes the number of function evaluations by only evaluating at the point that shows the maximum expected improvement (EI)~\cite{jones1998efficient}. The value of EI is estimated by a Gaussian process (GP) regressor~\cite{rasmussen2004gaussian}. At each iteration of EGO we want to find a latent code that is expected to best improve upon the current optimum and then evaluate at that point. Since it is also possible to have latent codes corresponding to invalid designs, we are dealing with a constrained SBO problem. Inspired by Refs.~\cite{basudhar2012constrained} and \cite{gelbart2014bayesian}, we solve the following constrained optimization problem at each iteration:
\begin{equation}
\begin{aligned}
\max_{\bm{c}} ~~~& \text{EI}(\bm{c})\text{Pr}(\mathcal{C}(\bm{c})) \\
\text{s.t.} ~~~& \text{Pr}(\mathcal{C}(\bm{c})) \geq 0.5
\label{eq:ei_constrained}
\end{aligned}
\end{equation}
where $\mathcal{C}(\bm{c})$ is an indicator of feasibility at $\bm{c}$ (\ie, whether the constraints are satisfied or whether the objective function has definition). In practice, we use a GP classifier~\cite{rasmussen2004gaussian} to estimate $\text{Pr}(\mathcal{C}(\bm{c}))$, where $\mathcal{C}(\bm{c})=0$ indicates the shape corresponds to $\bm{c}$ is self-intersecting or the simulation is not successful. At each iteration $t$, we find the solution $\bm{c}^{(t)}$ to Eq.~(\ref{eq:ei_constrained}) and evaluate at $\bm{c}^{(t)}$. The hyperparameters of both the GP regressor and the GP classifier are optimized by maximizing the log marginal likelihood (LML). We direct interested readers to~\cite{rasmussen2004gaussian} for details of implementing a GP regressor/classifier and using LML for hyperparameter optimization.

The last iteration of the first-stage optimization provides a latent code $\bm{c}^\dagger$ that roughly locks down the major features of an optimal design. The true optimal latent code is likely to be near $\bm{c}^\dagger$. The \textit{second-stage optimization} then refines this solution by jointly optimizing $\bm{c}\in\mathbb{R}^d$ and $\bm{z}\in\mathbb{R}^{d'}$. We set $\bm{z}\sim P_{\bm{z}}=\mathcal{N}(\bm{\mu_z},\bm{\Sigma_z})$, which means to synthesize realistic designs we need to sample $\bm{z}$ near $\bm{\mu_z}$. Thus, in the second stage, we start from $(\bm{c}^\dagger, \bm{\mu_z})$ and use GA to search for a refined solution.

\section{Experiment: Airfoil Synthesis and Shape Optimization}
\label{sec:experiment}

In this section, we test the performance of the B\'ezier-GAN as a parameterization in two aspects: (1) \textit{representation capacity}, or the ability to cover the design space, and (2) \textit{representation compactness}, or the ability of using the least number of parameters to cover a sufficient design space while every point in the parametric space maps to a valid design.

\subsection{Dataset and Preprocessing}

We use the UIUC airfoil database\footnote{\url{http://m-selig.ae.illinois.edu/ads/coord_database.html}} as our training data for the B\'ezier-GAN. It provides the geometries of approximately 1600 real-world airfoil designs that cover applications from low Reynolds number airfoils for UAVs and model aircraft to jet transports and wind turbines. Each design is represented by discrete 2D coordinates along their upper and lower surfaces. From the dataset, we removed outliers with unrealistic appearance. The number of coordinates for each airfoil is inconsistent across the database, so we use B-spline interpolation to obtain consistent shape representations. Specifically, we interpolate 192 points over each airfoil with the concentration of these points along the B-spline curve based on the curvature~\cite{je2001optimized}. The preprocessed data are visualized at the top of Fig.~\ref{fig:airfoil}. We have published the preprocessed data and the code for reproducing the experimental results\footnote{\url{https://github.com/IDEALLab/bezier-gan}}.

\subsection{B\'ezier-GAN Parameterization}
\label{sec:resuls_parameterization}

The latent codes and the noise were concatenated and fed into the generator. The generator has two branches~\textemdash~one generates control points and weights through dense layers and deconvolutional layers~\cite{zeiler2010deconvolutional} and the other generates parameter variables via only dense layers. The B\'ezier layer combines the outputs of the two branches and synthesizes 2D point coordinates along the surface curve. The discriminator takes in the coordinates and predicts the source of the input and the latent codes with dense layers and convolutioinal layers. Batch normalization and Leaky ReLU activation were used at each intermediate layer. The number of training steps was 10,000 and the batch size was 32. We set $P_{\bm{c}}=\text{Unif}(\bm{0},\bm{1})$ and $P_{\bm{z}}=\mathcal{N}(\bm{0},0.5\bm{I})$. The detailed implementation of B\'ezier-GAN can be found in our code. For the loss function shown in Eq.~(\ref{eq:loss}), we used $\lambda_0=1$ and $\lambda_1=\lambda_2=\lambda_3=\lambda_4=10$. We trained the B\'ezier-GAN using an Adam optimizer~\cite{kingma2014adam} on a Nvidia Titan X GPU. The wall-clock training time is about 1 hour, and the inference takes less than 15 seconds.

\begin{figure}[hbt!]
\centering
\includegraphics[width=1\textwidth]{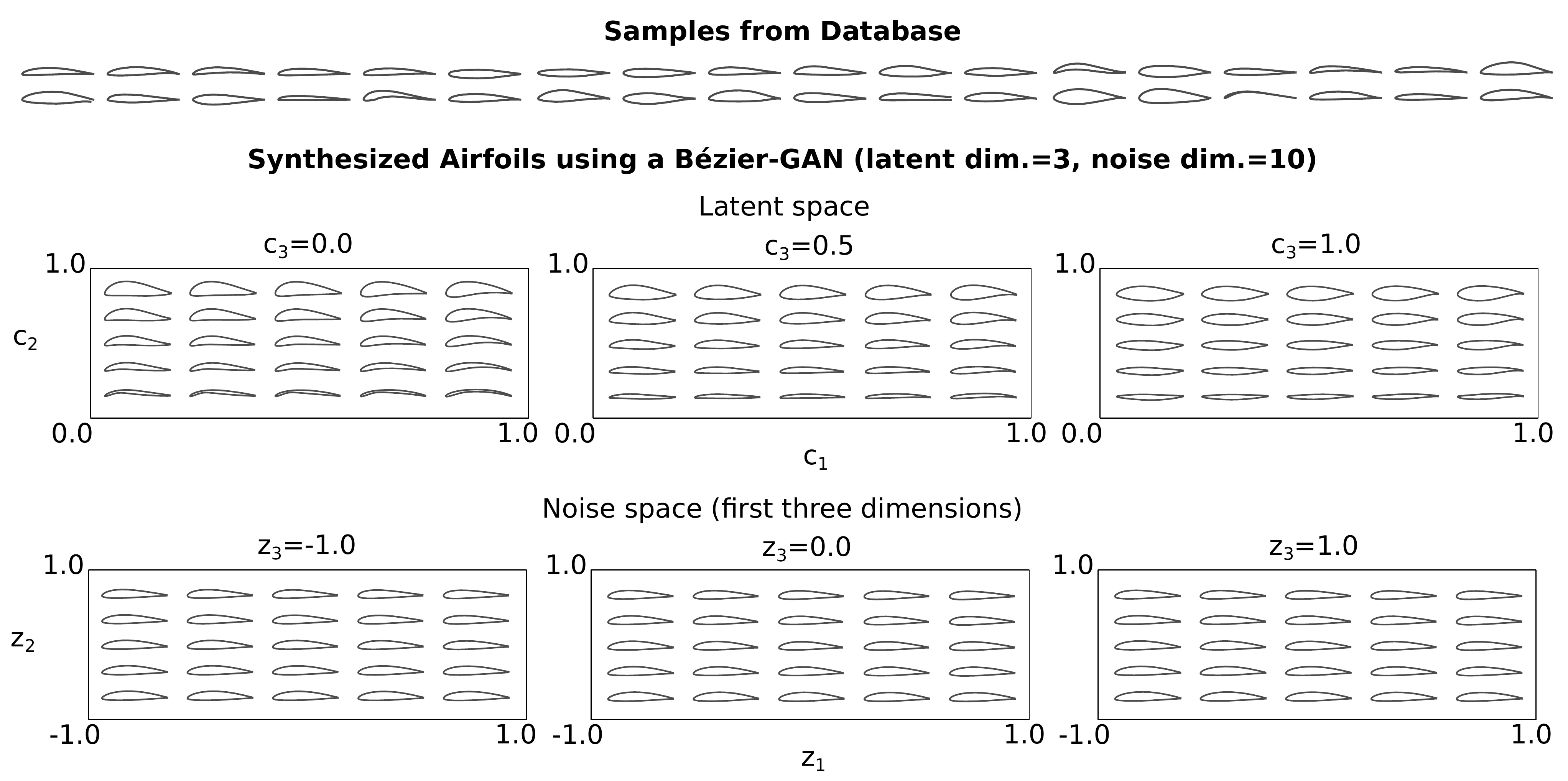}
\caption{Examples in the airfoil database and synthesized airfoil shapes in the latent space and the noise space (visualized by uniform slices of multiple two-dimensional spaces).}
\label{fig:airfoil}
\end{figure}

Figure~\ref{fig:airfoil} shows the synthesized airfoils by linearly interpolating points in the latent space and the noise space using a trained B\'ezier-GAN. The three-dimensional latent space captured large shape variations with respect to features such as thickness and camber/cord line curvature. Those are major features of the airfoil geometry. In contrast, shapes in the noise space only show small variations when fixing the latent codes. This indicates that the noise space captured minor features.

\begin{figure}[hbt!]
\centering
\includegraphics[width=.8\textwidth]{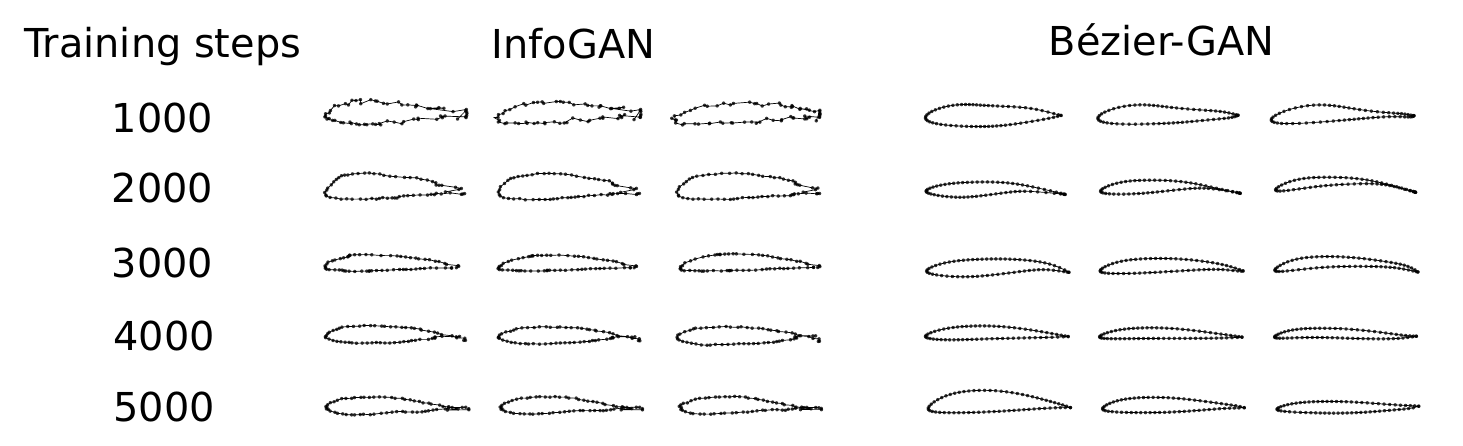}
\caption{Synthesized shapes during the training processes for an InfoGAN (left) and a B\'ezier-GAN (right).}
\label{fig:iterations}
\end{figure}

Figure~\ref{fig:iterations} compares synthesized shapes during the training processes of a B\'ezier-GAN and an InfoGAN. It demonstrates that the B\'ezier-GAN converged smooth and realistic airfoil shapes in far fewer training samples compared to the InfoGAN. 
 
\begin{figure}[hbt!]
\centering
\includegraphics[width=1\textwidth]{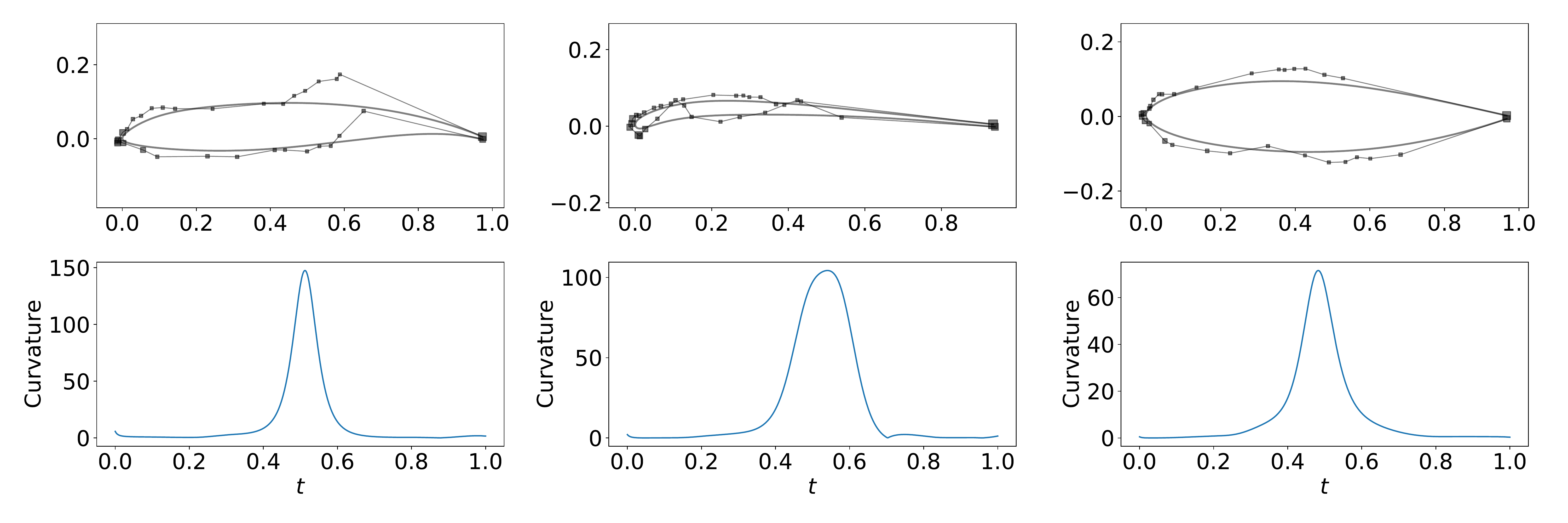}
\caption{Control points (square markers with sizes indicating the magnitudes of weights $w$) and curvatures of three randomly synthesized airfoils.}
\label{fig:curvature}
\end{figure}

One other advantage of the B\'ezier-GAN over an InfoGAN or other discrete parameterizations like SVD~\cite{poole2019efficient} is that it synthesizes continuous B\'ezier curves through the B\'ezier layer, rather than directly generating discrete surface point representations. This means that the resultant shapes will always have continuous curvature. This guarantee benefits aerodynamic performance, since past work has shown that aerodynamic performance is strongly dependent on the shape's curvature continuity~\cite{korakianitis2012aerodynamic,song2014numerical,shen2016experimental}.
As Figure~\ref{fig:curvature} shows, the synthesized airfoils exhibit smooth curvature profiles. We compute the curvature at a surface point $(x(t), y(t))$ via:
\begin{equation}
\kappa(t) = \frac{\dot{x}\ddot{y}- \ddot{x}\dot{y}}{(\dot{x}^2 + \dot{y}^2)^{\frac{3}{2}}},
\end{equation}
where $x$, $y$, and their derivatives can be obtained from Eq.~(\ref{eq:bezier}), given control points $P_1,...,P_n$ and weights $w_1,...,w_n$ as the generator's intermediate layer output. We also visualize the control points and the weights in Fig.~\ref{fig:curvature}. If one needs to constrain the curvature (\eg, to improve aerodynamic performance, reduce simulation complexity, or satisfy manufacturing tolerances), we can add another regularization term which contains $\kappa(t)$ to B\'ezier-GAN's loss function in Eq.~(\ref{eq:loss}). However, this is beyond the scope of the current work and we leave it for future study.

\begin{figure}[hbt!]
\centering
\includegraphics[width=0.4\textwidth]{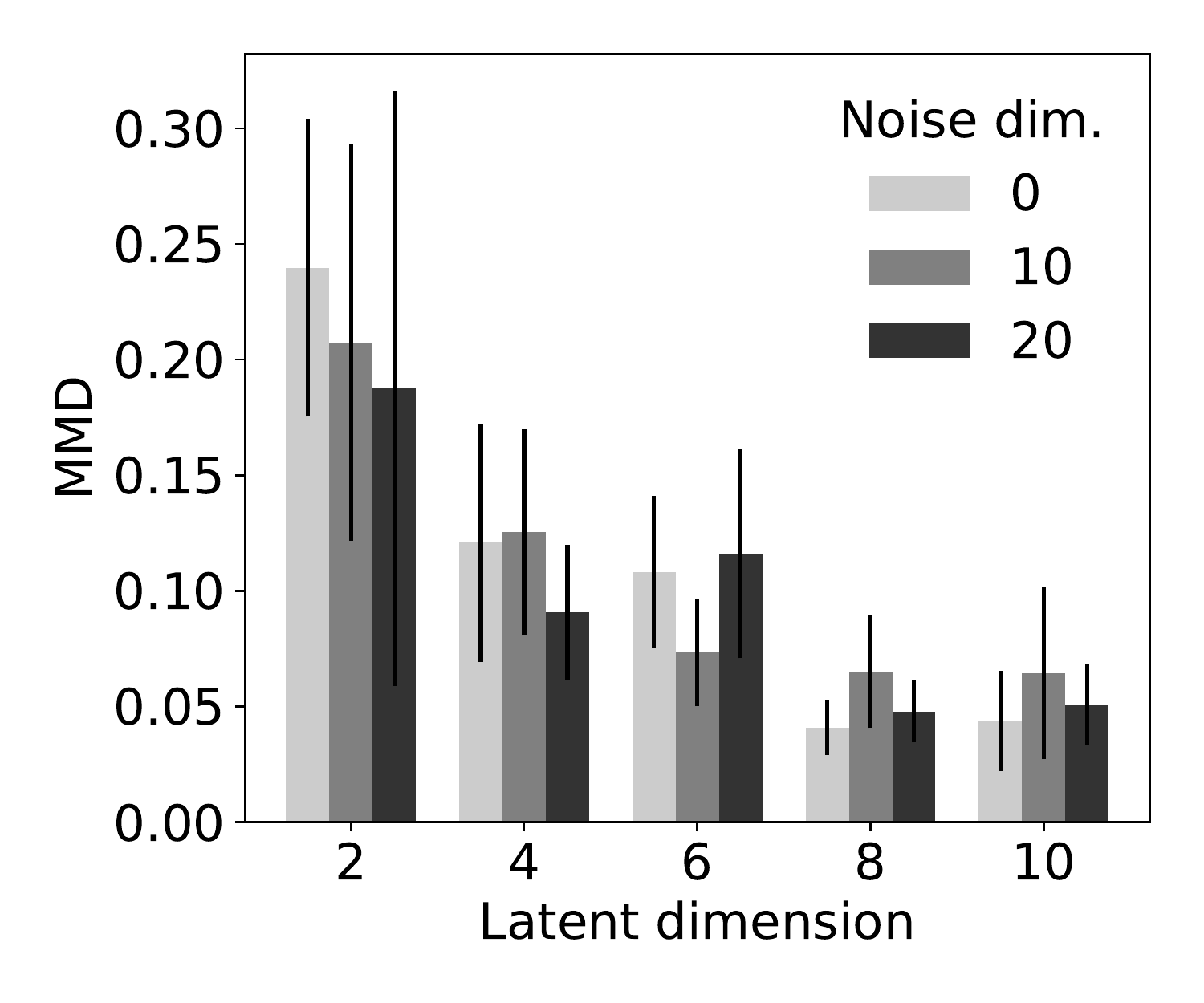}
\caption{Maximum mean discrepancy (MMD) metrics of trained B\'ezier-GANs. The error bars indicate 95\% confidence intervals.}
\label{fig:mmd}
\end{figure}

To study the effects of latent/noise dimensions on the B\'ezier-GAN parameterization and the optimization performance, we trained multiple B\'ezier-GANs with different combinations of latent dimension ($d=2,4,6,8,10$) and noise dimension ($d'=0,10,20$). We used kernel maximum mean discrepancy (MMD)~\cite{gretton2012kernel} to evaluate the quality of generated designs. The MMD metric measures how well our generator approximates the real data distribution:
\begin{equation}
\text{MMD}^2(P_{data}, P_G) = \\ \mathbb{E}_{\bm{x}_d,\bm{x}'_d\sim P_{data};\bm{x}_g,\bm{x}'_g\sim P_G}\left[k(\bm{x}_d,\bm{x}'_d) - 2k(\bm{x}_d,\bm{x}_g) + k(\bm{x}_g,\bm{x}'_g)\right],
\end{equation}
where $k(\bm{x},\bm{x}') = \exp{\left(-\|\bm{x}-\bm{x}'\|^2/(2\sigma^2)\right)}$ is a Gaussian kernel and $\sigma$ is the kernel length scale which was set to 1.0. A lower MMD indicates that the generator is better at generating realistic designs. Results shown in Fig.~\ref{fig:mmd} were collected via ten runs for each latent and noise dimension configuration. It indicates that the latent dimension plays a key role in improving the generator. Also, no significant improvement is shown after the latent dimension reaches eight, indicating that major features of the airfoil geometry can be adequately encoded with eight latent variables. Ideally, one would expect lower MMD values when the noise dimension is larger, as the noise can encode minor features that are not captured by latent codes. However, in our results that was true only when the latent dimension was two. One explanation for this is because increasing the number of latent dimensions leaves less room for the noise variables to control meaningful shape variation, as the latent codes have a higher priority of capturing shape variation than the noise variables. Thus, the noise variables are less effective when the latent dimension is high.

\begin{figure}[hbt!]
\centering
\includegraphics[width=0.95\textwidth]{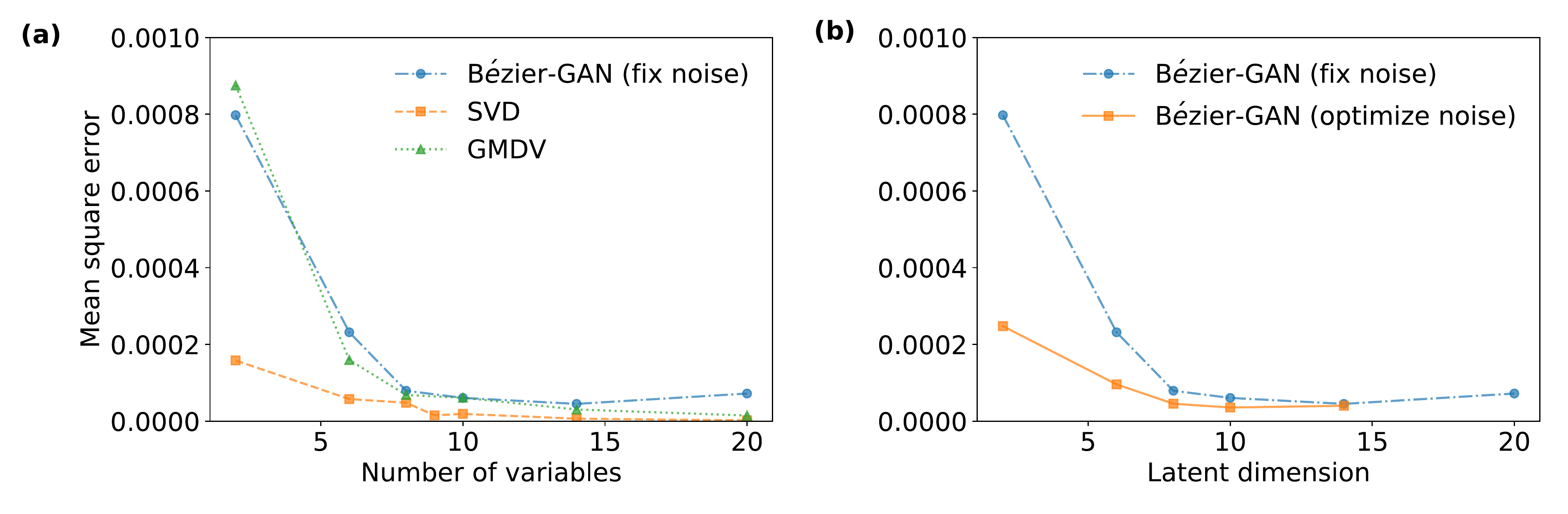}
\caption{The fitting test results showing the performance of recovering the UIUC airfoils (Note that the fitting performance is biased towards SVD and GMDV and will not solely depend on the design space coverage, since both SVD and GMDV have analytical solutions to the least squares problem whereas B\'ezier-GAN's solution is approximate).}
\label{fig:fitting_errors}
\end{figure}

We benchmark B\'ezier-GAN against three state-of-the-art parameterization approaches, namely GMDV~\cite{kedward2020towards}, SVD~\cite{poole2019efficient}, and FFD~\cite{sederberg1986free,masters2017geometric}. We also perform a fitting test to evaluate a parameterizaton's ability to recover a wide range of existing airfoil designs. Specifically, we perform least squares fitting to match the synthesized airfoils with the UIUC airfoils under different parameterizations and numbers of design variables. The results are shown in Fig.~\ref{fig:fitting_errors} (we set B\'ezier-GAN's noise dimension to 10). Lower mean square error (MSE) indicates better coverage of the UIUC data. Note that the fitting performance is biased towards SVD and GMDV and will not solely depend on the design space coverage, since both SVD and GMDV have analytical solutions to the least squares problem whereas B\'ezier-GAN's solution is approximate. This approximation is because the fitting problem is non-analytical and non-convex for B\'ezier-GAN, requiring iterative methods to find the (possibly sub-optimal) least squares fit. While this hinders B\'ezier-GAN fitting performance on the training data, it will not necessarily affect its performance in design optimization, as we show in the next section. The fitting results show that all the tested methods converge to a plateau as the number of variables increases. As shown in Fig.~\ref{fig:fitting_errors}a, SVD has the lowest MSE, which is reasonable as it uses UIUC airfoils as training data and we can obtain the exact optima of the least squares problem. B\'ezier-GAN with fixed noise has a result similar to GMDV. In Fig.~\ref{fig:fitting_errors}b, we show two scenarios: (1) we only optimize latent codes and fix noise variables during the least squares fitting; and (2) we optimize both latent and noise variables. The figure shows that as the latent dimension increases, optimizing noise variables contributes less to lowering the MSE (\ie, has less control on shape variation). This result offers one indication or test as to when performing the two-stage optimization is beneficial.

The elbow point in Fig.~\ref{fig:fitting_errors} (\ie, the number of design variables where the corresponding fitting error drops with relatively small difference) can indicate the lowest number of variables needed to reasonably cover the original design space. For B\'ezier-GAN, the elbow point suggest a latent dimension of 8. For SVD and GMDV, their elbow points locate at 8 and 9, respectively. Thus, we will set the number of modes for SVD and GMDV according to these numbers in the rest of our experiments.

\begin{figure}[hbt!]
\centering
\includegraphics[width=0.75\textwidth]{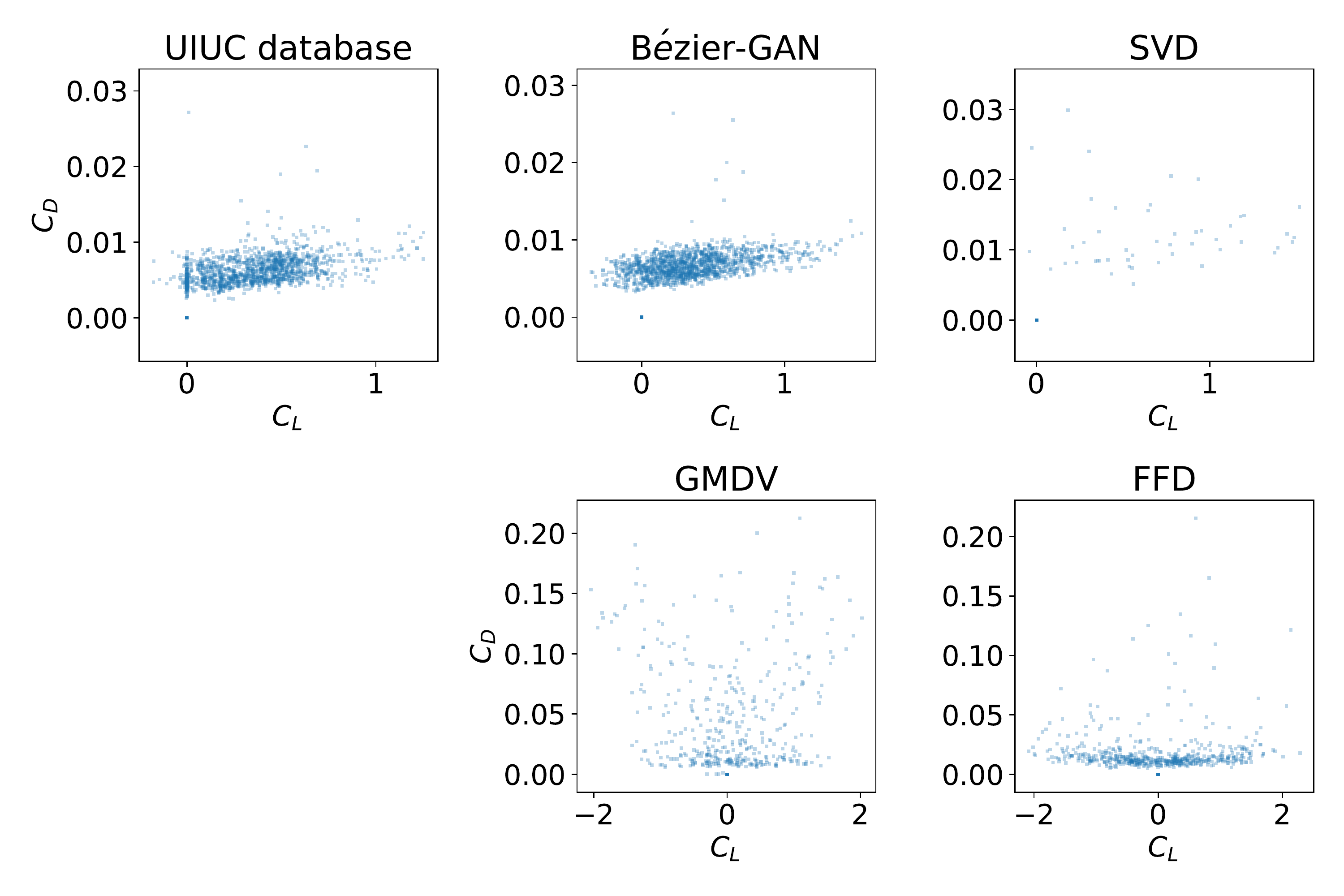}
\caption{Comparison of parameterizations' performance spaces under random samples. Points at the origin represent performances unable to compute due to self-intersecting shapes or simulation failures, which we call \textit{invalid performance}. Note that the plots are in different scales.}
\label{fig:clcd}
\end{figure}

In addition to the design space coverage study, we also compared the performance space coverage (\ie, how the synthesized airfoils cover the space of $C_L$ and $C_D$) of B\'ezer-GAN, SVD, GMDV, and FFD (Fig.~\ref{fig:clcd}). For FFD, we use a set of $3\times 4$ control points and fix their $x$ coordinates based on~\cite{masters2017geometric}. It shows that the performance distribution of B\'ezier-GAN airfoils best matches the UIUC database. SVD's performance coverage is similar to the data, but contains a large portion of invalid performances, where shapes are self-intersecting or cause unsuccessful simulations. Both GMDV and FFD have larger coverage of the performance space compared to the input data. This is expected since GMDV and FFD are not explicitly trained on the input data and thus their parameterizations readily sample designs far beyond the original design space. 

\begin{figure}[hbt!]
\centering
\includegraphics[width=0.8\textwidth]{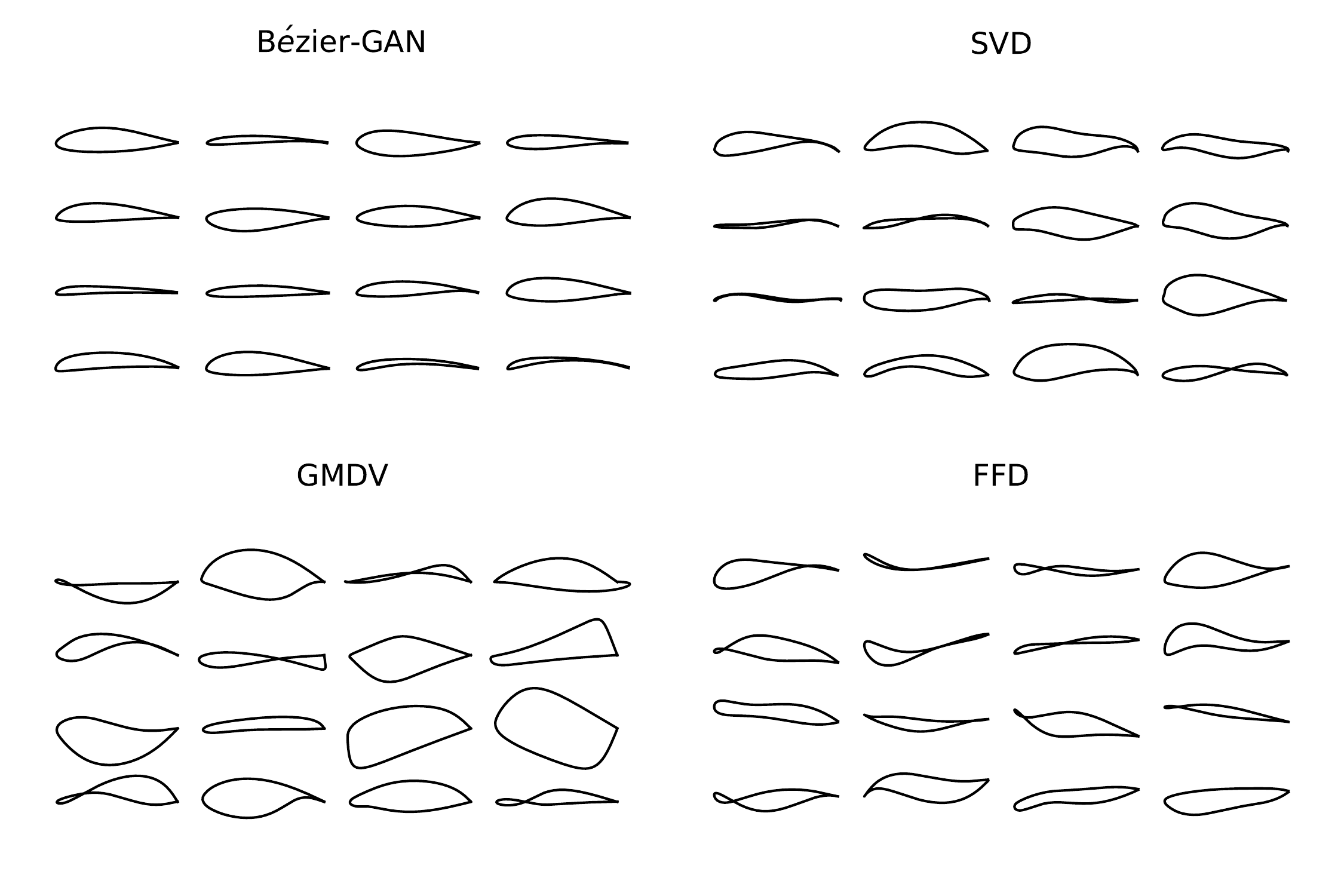}
\caption{Random airfoils synthesized by parameterizations.}
\label{fig:samples}
\end{figure}

The coverage of the design space and the performance space indicate the level of representation capacity. Another crucial property of a parameterization is its representation compactness, which indicates the proportion of useful designs in its design space. Figure~\ref{fig:samples} shows airfoils synthesized by randomly sampling points under different parameterizations (\ie, random samples of the design space). 
Specifically, the B\'ezier-GAN airfoils are synthesized by using latent codes and noise variables drawn from their respective distributions (mentioned in Section~\ref{sec:resuls_parameterization}). Knowing the distribution of design variables corresponding to valid designs (\textit{valid design variables}) can benefit design space exploration. For the other three parameterizations where the distribution of valid design variables is unknown, the airfoils are synthesized by design variables drawn uniformly at random within specific bounds. For GMDV, we set the bounds according to Ref.~\cite{kedward2020towards} (more details can be found in the code). For SVD, we set the bounds as the minimum bounding box of design variables corresponding to the database. For FFD, we set the bounds to be $\pm 0.2$ perturbation of the NACA~0012 airfoil.
All designs synthesized by the B\'ezier-GAN look realistic, while other parameterizations contain a larger proportion of invalid (\eg, self-intersected) designs. This indicates that B\'ezier-GAN has a higher representation compactness than the other parameterizations, which is expected as B\'ezier-GAN's objective forces each point in the latent or noise space to map to a design resembling the training data.

\subsection{Optimization}

In Section~\ref{sec:resuls_parameterization}, we show the representation capacity of B\'ezier-GAN through the coverage of the design space and the performance space. We also visualize the design space through randomly synthesized designs, which provides an indication for representation compactness. But why do either of these properties matter? In this section, we demonstrate the benefits of such properties for accelerating an aerodynamic optimization task. The performance of global optimization is closely related to the parameterization's representation capacity and compactness. The low representation capacity may lead to a sub-optimal solution, since a better one cannot be represented by the parameterization. Meanwhile, low representation compactness can contribute to slow convergence due to the curse of dimensionality and the vast amount of invalid designs an optimizer might evaluate; this wastes the evaluation budget.

The below optimization task is only a means to assess the impact of each parameterization's representation capacity and compactness. To make the below experiments easy and fast to replicate and evaluate by other researchers, we use XFOIL~\cite{drela1989xfoil} to compute the lift and drag coefficients $C_L$ and $C_D$ of candidate airfoils. However, our approach is not limited to XFOIL. One can apply our techniques to any CFD or performance code including Reynolds-Averaged Navier-Stokes simulations or LES. The specific choice of CFD simulation method we use below is not central to evaluating the key contributions of the paper, though evaluating the impact of better parameterizations on more expensive and advanced flow simulations is an interesting area for future work.

Specifically, for the following experiments our optimization objective is to maximize the lift to drag ratio, \ie, $f(\bm{x})=C_L/C_D$. The operating conditions are set as follows: Reynolds number $Re = 1.8 \times 10^6$, Mach number $Ma = 0.01$, and angle of attack $\alpha = \ang{0}$.

We trained ten B\'ezier-GANs for each configuration of latent/noise dimensions. For each configuration, we obtained one optimization result by using each of the ten B\'ezier-GANs as the parametric model and studied the statistics (\eg, mean and standard deviation of the optimization history) of those results. Thus, different from other parameterizations, the variance in B\'ezier-GAN's optimization results also depends on the stochasticity of trained B\'ezier-GAN models. At the first-stage optimization, each latent code was bounded in $[-0.2, 1.2]$. After getting the near-optimal latent code $\bm{c}^\dagger$, we perform the second-stage optimization by bounding each latent code in $[c^\dagger_i-0.1, c^\dagger_i+0.1], i=1,...,d$ and each noise variable in $[-1.0, 1.0]$.

\begin{figure}[hbt!]
\centering
\includegraphics[width=0.7\textwidth]{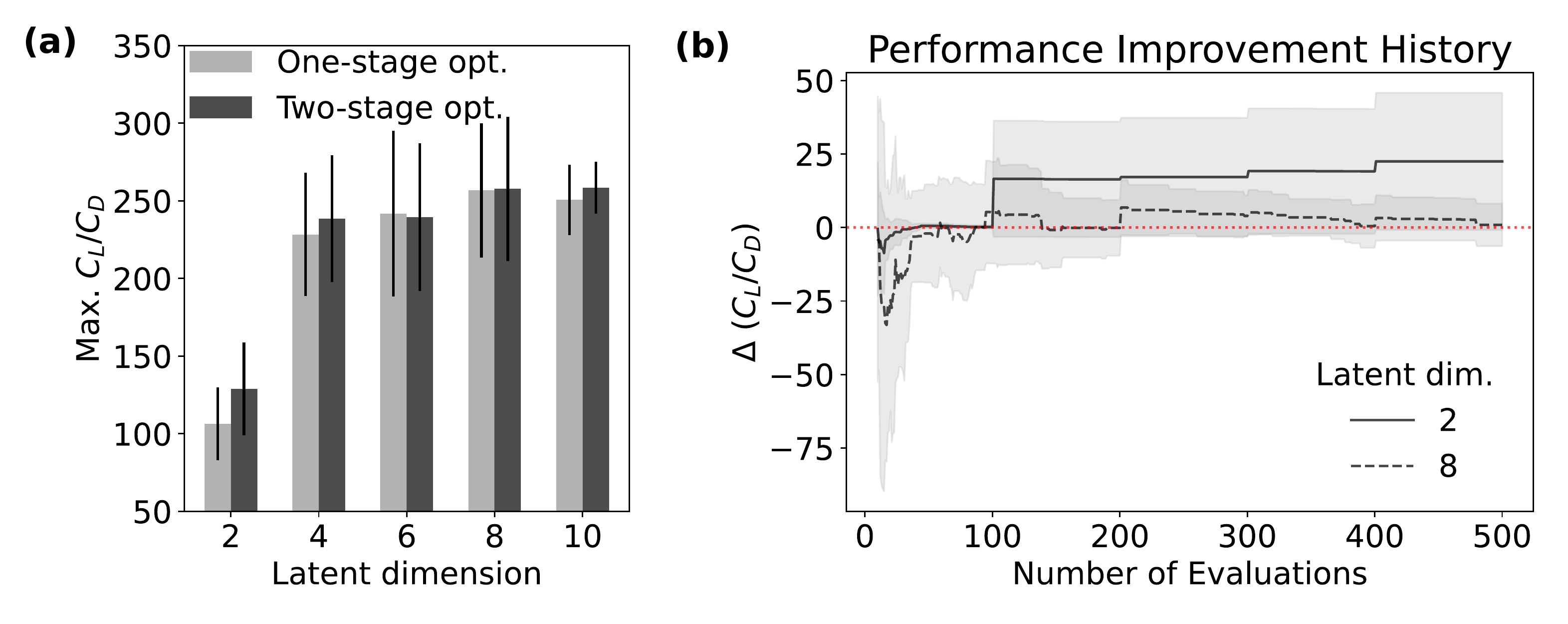}
\caption{Optimization results for B\'ezier-GAN parameterization with (\ie, two-stage optimization or TSO) and without (\ie, one-stage optimization or OSO) refining the noise variables. The noise dimensions was set to 10. Plot (a) shows the final optimal $C_L/C_D$ over different latent dimensions. Plot (b) shows the performance improvement history of TSO over OSO ($\Delta(C_L/C_D)$). The dotted horizontal line represents $\Delta(C_L/C_D)=0$, referring to the case where TSO has exactly the same performance as OSO.}
\label{fig:opt_refine}
\end{figure}

Figure~\ref{fig:opt_refine} compares the optimization performance under the two-stage optimization (TSO) and the one-stage optimization\footnote{We used EGO for the one-stage optimization.} (OSO, \ie, the case where we assign the overall budget to only optimizing the latent codes). Figure~\ref{fig:opt_refine}a shows that compared to OSO, TSO is more effective in terms of the final optimal $C_L/C_D$ when the latent dimension is small. Figure~\ref{fig:opt_refine}b shows the $C_L/C_D$ improvement of TSO over OSO ($\Delta(C_L/C_D)$) is much more significant throughout the optimization history when the latent dimension is lower. These results make sense because, as already indicated by the MMD values (Fig.~\ref{fig:mmd}) and the fitting errors (Fig.~\ref{fig:fitting_errors}b), the increase of the latent dimension leaves less room for the noise variables to control shape variation, and hence optimizing noise variables at the second stage becomes less useful.

\begin{figure}[hbt!]
\centering
\includegraphics[width=0.6\textwidth]{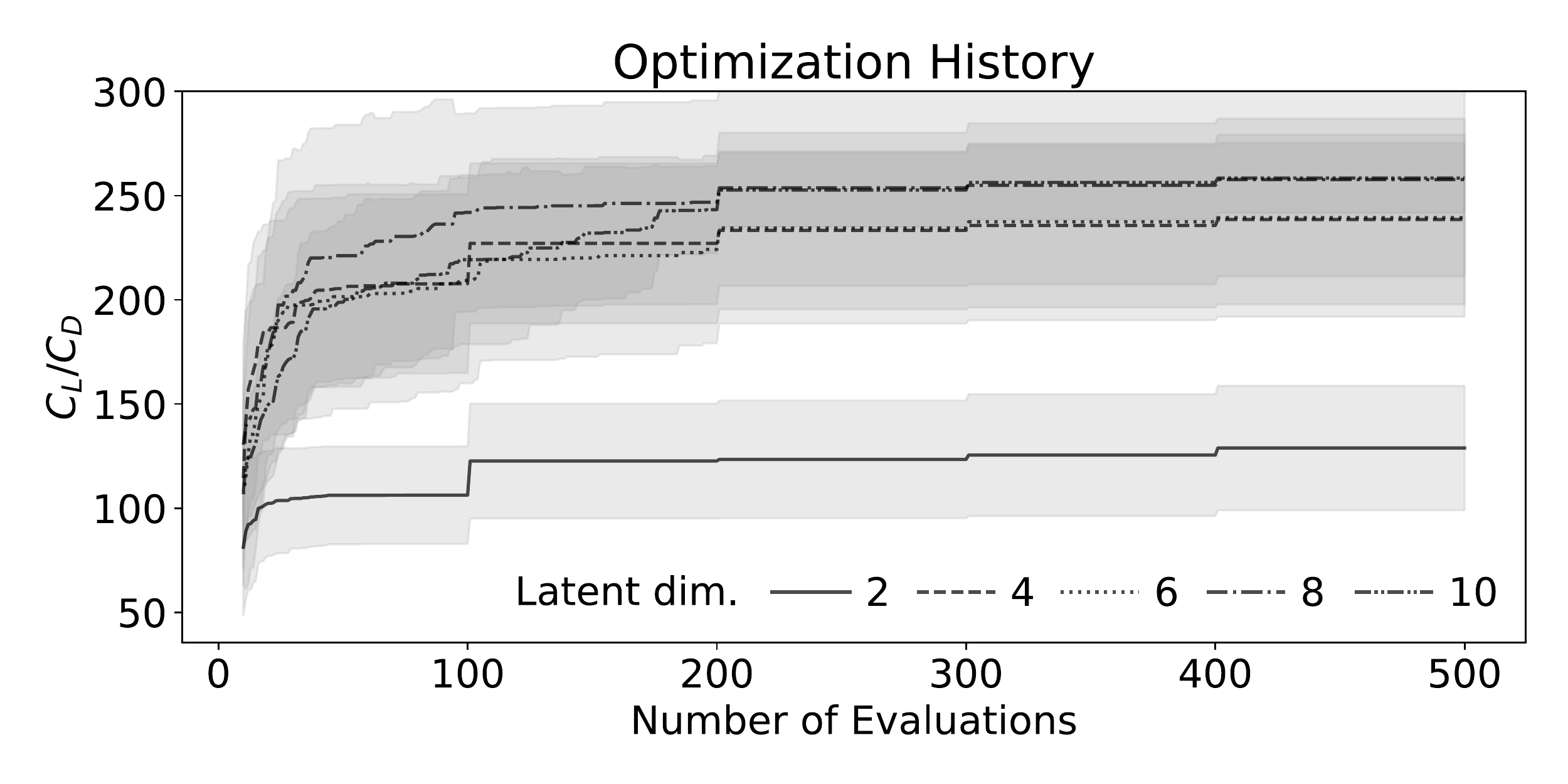}
\caption{Optimization history under B\'ezier-GAN parameterization with different latent dimensions. The noise dimension was fixed to 10.}
\label{fig:opt_history_latent}
\end{figure}

The optimization performance is affected by both a parameterization's representation capacity and compactness. While low representation capacity may lead to solutions with limited performance, low representation compactness may explore larger regions of invalid designs and hence be slower at finding an optimum. To some extent, we can evaluate a parameterization by comparing its corresponding optimization history. As shown in Fig.~\ref{fig:opt_history_latent}, the optimization performance improved when increasing the latent dimension from two to eight, but further increases in the latent dimension decreases the convergence speed. This result is consistent with the aforementioned MMD values (Fig.~\ref{fig:mmd}) and fitting errors (Fig.~\ref{fig:fitting_errors}). All our results indicate that a latent dimension of eight suffices with respect to the representation capacity for our specific optimization problem.

\begin{figure}[hbt!]
\centering
\includegraphics[width=0.6\textwidth]{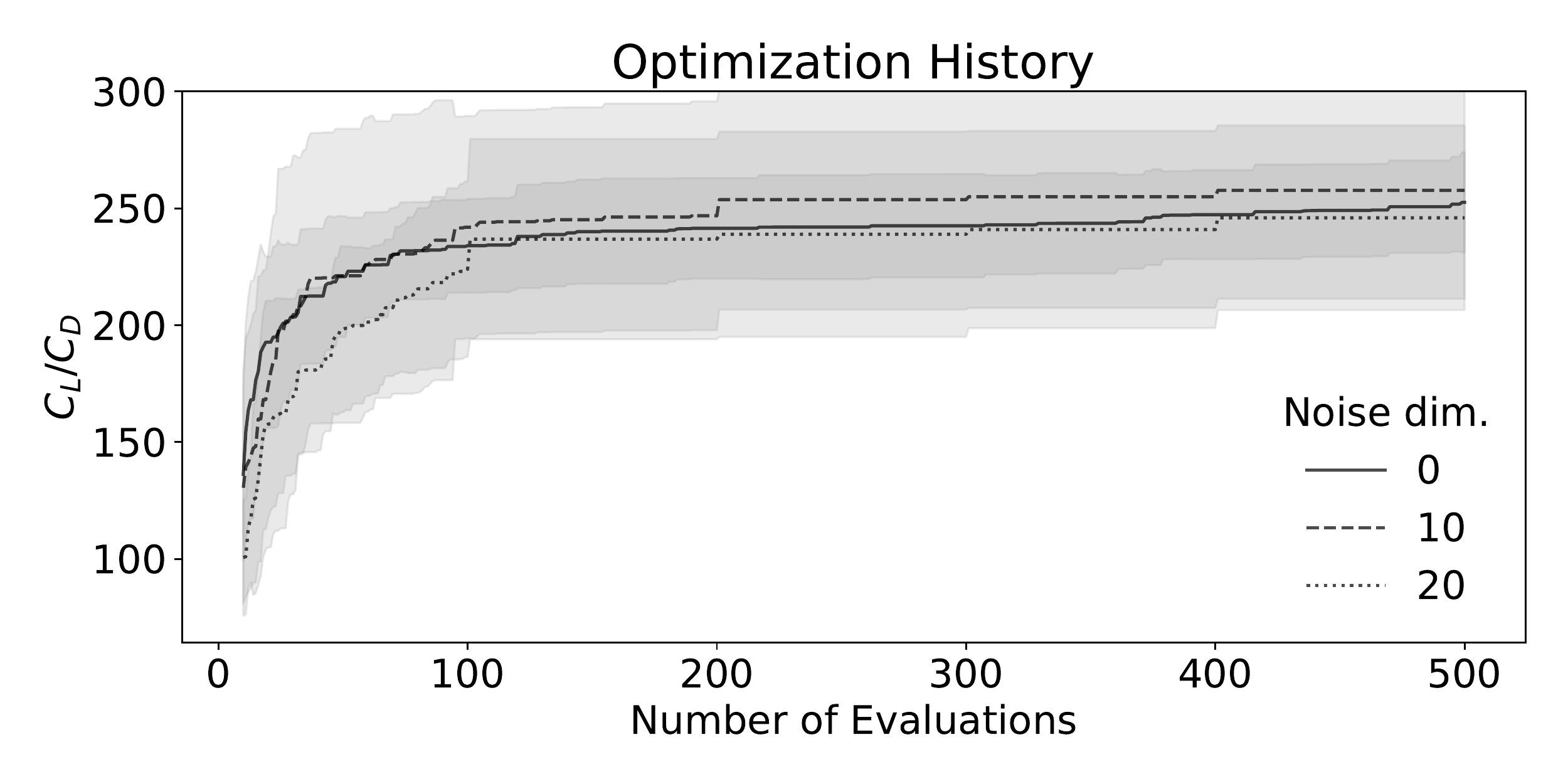}
\caption{Optimization history under B\'ezier-GAN parameterization with different noise dimensions. The latent dimension was fixed to 8.}
\label{fig:opt_history_noise}
\end{figure}

Figure~\ref{fig:opt_history_noise} shows that although the noise dimension cannot affect the optimization performance as significantly as the latent dimension does, having noise variables is still better than not. Note that as the noise dimension goes up, the representation capability will go up (which leads to an improved final optimal solution under the noise dimension of 10 over 0), but the optimization convergence may slow due to a reduced representation compactness and the curse of dimensionality, \eg, using a noise dimension of 20 coverged slower than when using 10).

\begin{figure}[hbt!]
\centering
\includegraphics[width=0.6\textwidth]{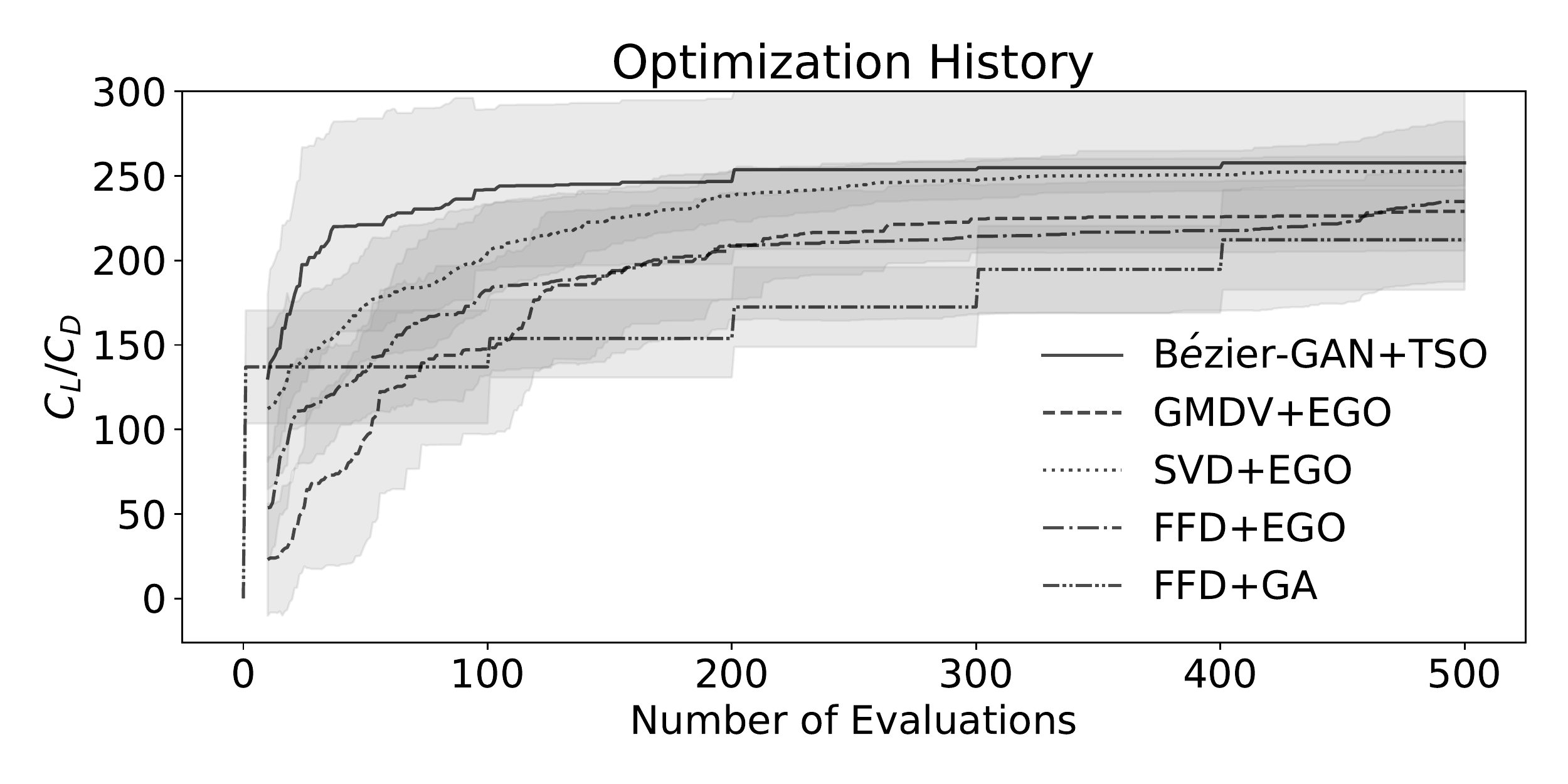}
\caption{Optimization history under B\'ezier-GAN and other parameterizations. The performance reached by Bezier-GAN+TSO in 200 evaluations takes the next best method 500 evaluations to reach.}
\label{fig:opt_history_conventional}
\end{figure}

We also compared our method to several state-of-the-art parameterization methods (Fig.~\ref{fig:opt_history_conventional}). The latent and noise dimensions of B\'ezier-GAN were set to 8 and 10, respectively. For GMDV and SVD, we set the number of design variables to 8 and 9, respectively, according to their elbow points in the fitting test results (Fig.~\ref{fig:fitting_errors}). We used $3\times 4$ control points for FFD, according to Ref.~\cite{masters2017geometric}. The design variable bounds for GMDV were set according to Ref.~\cite{kedward2020towards}. For SVD, we set the design variable bounds as the minimum bounding box of design variables corresponding to the database. For FFD, the design variable bounds were $\pm 0.2$ perturbation of the NACA~0012 airfoil. When performing GA, the population size was 100, and the chance of mutation (\ie, the probability of mutating an individual's parameter) was 0.1. In each generation, we chose the 30 best and 10 random individuals for crossover, and produced 5 children for each pair. We direct interested readers to our code for further implementation specifics.
Besides the aforementioned parameterizations, we also tested the optimization performances of nonuniform rational B-splines (NURBS)~\cite{lepine2000wing} and PARSEC~\cite{li1998manual}. However, we do not show their results in the figure as their performances were strictly worse than the SVD, GMDV, FFD, and B\'ezier-GAN parameterizations. We direct interested readers to Ref.~\cite{chen2019aerodynamic} for these results.
Each optimization history was averaged over ten runs. 

\begin{figure}[hbt!]
\centering
\includegraphics[width=1\textwidth]{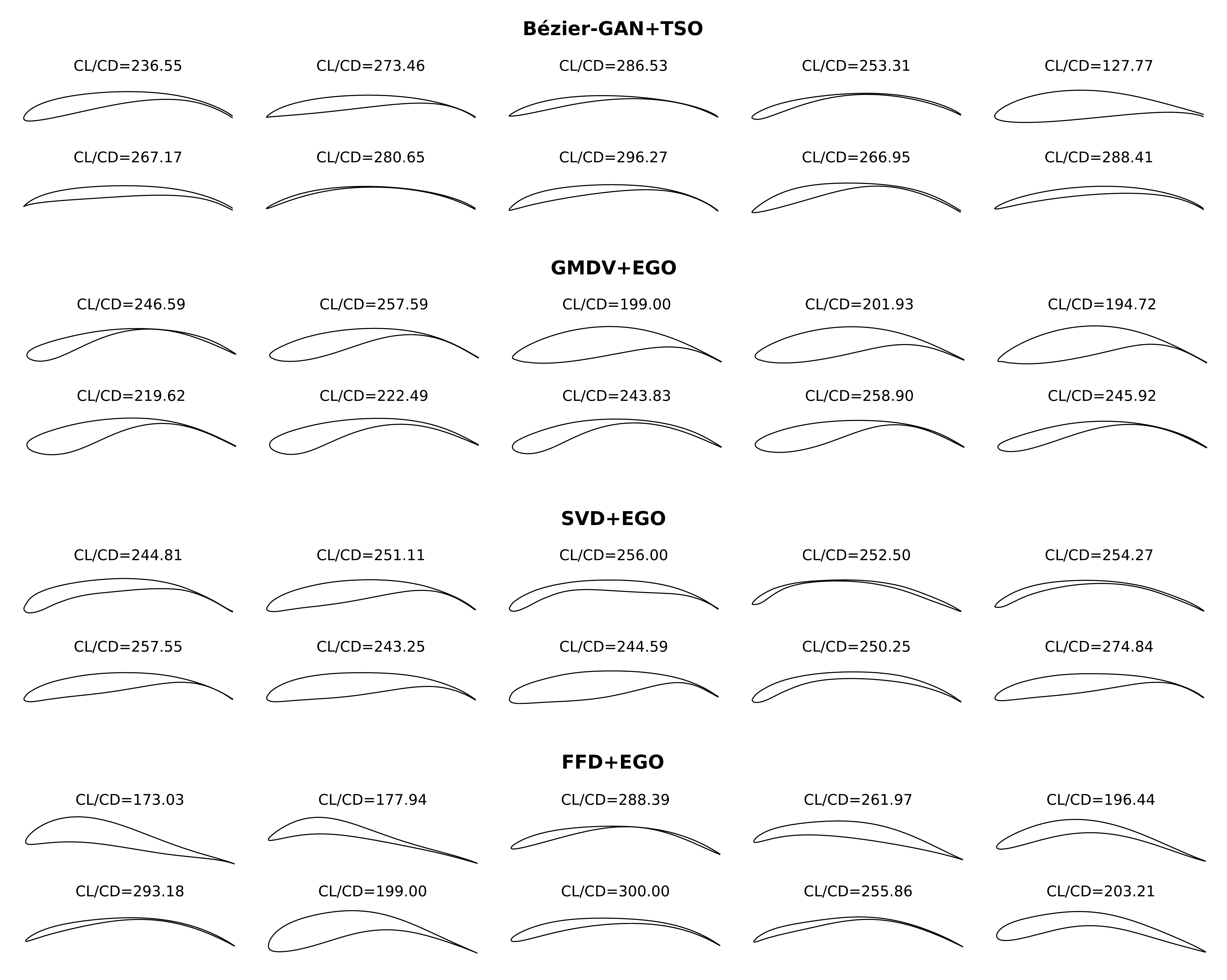}
\caption{Optimal airfoils for different parameterizations.}
\label{fig:opt_airfoils}
\end{figure}

The results show that the average $C_L/C_D$ value reached in 200 evaluations by the B\'ezier-GAN takes other methods at least 500 evaluations to reach. It indicates that the B\'ezier-GAN parameterization has much higher compactness so that fewer resources were wasted in exploring invalid and poor-performance regions. Figure~\ref{fig:opt_airfoils} shows the optimal airfoils for the different parameterizations. It is notable that different from other parameterizations, the variance of the B\'ezier-GAN results comes not only from the optimization process but also the trained B\'ezier-GAN model, since the training process is stochastic.

\section{Conclusion and Discussion}

We use the B\'ezier-GAN as a new parameterization for aerodynamic designs that possesses high representation compactness and sufficient representation capacity. The latent codes and noise variables encode major and minor shape variations, respectively. We then propose using this parameterization as part of a two-stage design optimization method. Our results show that the B\'ezier-GAN accelerates convergence and finds optimal designs with higher performance than those found by other state-of-the-art parameterization methods.

We believe that this phenomenon is induced by the representation capacity and compactness of parameterizations and is independent of the specific CFD approach used to evaluate the airfoil. That is, a space that better describes airfoil variation should help during optimization, even if such a space was not directly designed to aid in that task. We agree though that the simulation environment in this paper is simple and that we are admittedly assuming that one would see similar optimization improvements under different performance evaluators, such as more advanced CFD. Likewise, we only demonstrate this for 2D airfoils, and we have not explicitly tested this hypothesis on 3D surfaces or larger wing segments. We expect that, in 3D with even greater design freedom, representations with better compactness and coverage would become even more important. However, we have not tested this. We believe that it might be an excellent area for future research.

In general, having a parameterization that separates major and minor shape deformation is useful for design optimization. Parameters with larger impact on the geometry can have higher priority during optimization, such that major features, which are usually the primary cause of the performance change, can be determined first. Although in this paper we used a two-way partitioning of design parameters, in an ideal case one would have all parameters ordered with respect to their importance on the geometry. Some linear dimensionality reduction methods like SVD can learn ordered representations. Efforts have also been made to extend this capability to nonlinear methods~\cite{rippel2014learning}. Yet so far, to the best of our knowledge, there is no equivalent for deep generative models. Related to this, one would need a novel global optimization method to efficiently optimize the ordered parameters. Future research can fill in these gaps.

While here we only demonstrate the B\'ezier-GAN's capability to parameterize airfoils, we can also train this new generative model to synthesize other smooth geometries such as hydrodynamic shapes. 

In this paper, we used EGO and GA in the first-stage and second-stage optimization, respectively. There are other ways to improve the optimal solution while maintaining fast convergence. For example, the optimum obtained by our method can be used as a good start point for gradient-based optimization methods (\eg, as in Berguin \etal~\cite{berguin2015method}). For future research, we can concatenate a trained B\'ezier-GAN generator and an automatic differentiation solver to obtain the gradient of the performance with respect to each of the latent codes and noise variables directly. In this way, gradients can be propagated to the reduced representation and help solve gradient-based optimization problems on the compact reduced space.

Overall, we hope that this paper highlights the promising directions that learned geometric parameterizations can play in shape and design optimization more generally, and how future work in such techniques can complement traditional optimization methods used by the optimization community.

\section*{Funding Sources}

This work was supported by The Defense Advanced Research Projects Agency (grant numbers 16-63-YFA-FP-059 and HR00111820009). The views, opinions, and/or findings contained in this article are those of the author and should not be interpreted as representing the official views or policies, either expressed or implied, of the Defense Advanced Research Projects Agency or the Department of Defense.

\bibliography{references}

\end{document}